# *Real Schur flow* computations, helicity fastening effects and Bagua-pattern cyclones

Jian-Zhou Zhu (朱建州)[a]

*Su-Cheng Centre for Fundamental and Interdisciplinary Sciences, Gaochun, Nanjing 211316, China and*

*Life and Chinese Medicine Study Center, Gui-Lin Tang Lab., 47 Bayi Cun, 366025 Yong'an, Fujian, PR China*

A semi-analytical algorithm is developed for simulating flows with the velocity gradient uniformly of the real Schur form. Computations for both decaying and driven cases are performed, exhibiting basic results for general conception and testing the specific notion of 'helicity fastening flows', and, creating the Jiu-Gong/Ba-Gua (ditetragonal/octagonal) pattern of cyclones resembling northern circumpolar cluster of Jupiter.

Keywords: rotating flows, real Schur flow, reduced model, helical compressible turbulence, compressibility reduction

[a] Electronic mail: jz@sccfis.org

## I. INTRODUCTION

As shown in Fig. 1 for a prototypical flow of the Taylor-Green[1] fashion described by Eq. (17) below, a two-component-two-dimensional coupled with one-component-three-dimensional (2C2Dcw1C3D) velocity field, with the 'horizontal' components, $u_1$ and $u_2$, presenting columnar/two-dimensional (2D: independent of the third/vertical coordinate $x_3$) patterns and the 'vertical' component $u_3$ fully three-dimensional (3D) one, has the velocity gradient uniformly of the real Schur form[2,3], thus a *real Schur flow* (RSF). Such an RSF is of an interesting mathematical value and physical relevance (Ref. 3 and references therein). A better understanding of how such anisotropic structure is maintained dynamically may also shed light onto the opposite problem of isotropization in Navier-Stokes flows (NSF), which is of course fundamentally important, especially for turbulence. However, RSF, formally simplifying, say, the full Navier-Stokes flow (NSF), can still be very complicated, welcoming also systematic numerical investigations to explore it in detail. Actually, as we shall see, finding and designing algorithms for computing RSF by itself requires in-depth investigation of the dynamics, especially the fine structures.

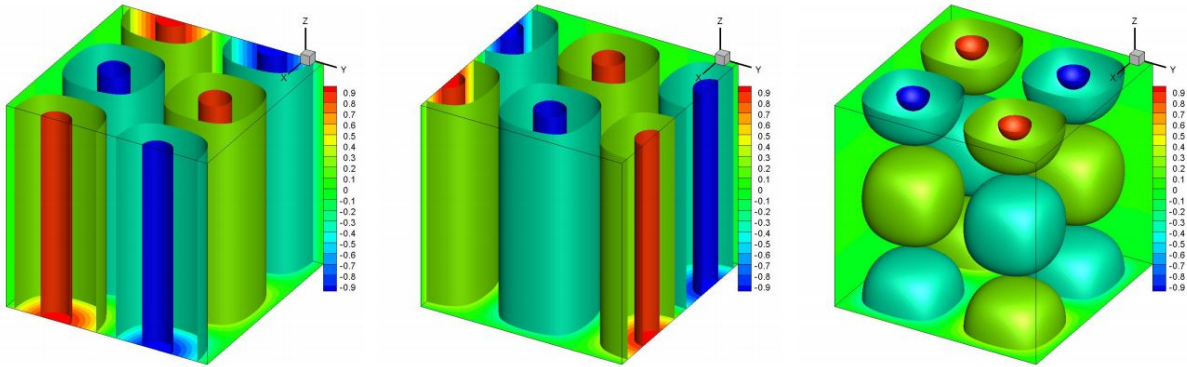

FIG. 1. The isosurfaces (of values $\pm 0.9$ and $\pm 0.3$) of the three components, $u_1$ (left), $u_2$ (middle) and $u_3$ (right), of the Taylor-Green-fashion 2C2Dcw1C3D velocity field.

Note that an RSF in the natural Navier-Stokes setting is of course unstable, but extra physical constraint, such as strong rotation or background magnetic field, among others, may stabilize it. Thus, singling it out for separate study makes a lot of sense. The classical incompressible Taylor-Proudman-Theorem states that 'everything' becomes two-dimensional, thus three-component-two-dimensional (3C2D) velocity, and the same argument without the condition of incompressibility leads formally to a 2C2Dcw1C3D RSF, with the additional 'horizontal' incompressibility[4]. The Taylor-Proudman theorem is of 'zeroth order', and multi-scale analysis is possible to give more systematic results, including those in even more complicated situation (e.g., Ref. 5 where it is shown that a reduced model on incompressible thermal convection can be deduced under the Taylor-Proudman constraint). RSF is also associated to the text-book fact[2] that any real matrix, such as the (local) velocity gradient matrix, can be transformed into the real Schur form with appropriate rotations of the coordinates. Thus, we may say that RSF is the intrinsic structure of a general flow, in a sense playing the role of local inertial frame in general relativity[3]. We have so far only studied the 'global' 2C2Dcw1C3D RSF with the velocity gradient uniformly of the general real Schur form, not the local matrix transformation, neither the associated kinematics. In other words, we do not need the transformation and we simply generalize the conventional 3C2D system to our 2C2Dcw1C3D RSF, to be closer to the full compressible NSF. However, it is interesting to remark that the studies, using the coordinate transformation with the objective to extract relevant information from the local real Schur form of the velocity gradient matrix to define or characterize some 'vortex' structures and other relevant issues, have actually been performed extensively.[6]

Numerical computation of RSF is nontrivial in the sense that the above defining property of the velocity does not directly specify the precise and complete form of the governing equation(s) appropriate for discretization and integration. Mathematically speaking, RSF lives in a subset of the (phase) space of NSF, requiring appropriate constraint(s) to avoid escaping. In other words, dynamically, a set of additional precise relations should be satisfied to have a self-consistent evolution of the system. Under the framework of NSF, it has been shown that such relations can be explicitly written down, in the sense of particular properties of the (primitive) variable itself as a function of space and time, for the inviscid case or for the slightly simplified viscosity model with, say, the dilatational effect neglected in the Stokesian model[3]. This makes it possible to define a dynamical projection operator straightforward for numerical implementation. For the more general cases, including models such as non-newtonian fluid or quantum NSF, the precise structures are typically not explicitly expressible but implicity specified, in the sense of some particular relation(s) between different (primitive) variables. It is possible that one may make use of (some of) the precise relations to apply in a 'brute-force' way the conventional dual-time stepping method[7] with the inner iterations (pseudo-time) for the accurate relaxation to the self-consistent state before advancing the physical time. How to ensure the convergence of the inner iteration and how accurate or how many (pseudo-time) steps are needed for the self-consistent RSF dynamics are in general not easy to be made clear and can be subtle. Another brute force is to resort to approaches related to molecular dynamics. For example, Gallis et al.[8] recently made a comparison between the direct simulation Monte Carlo (DSMC) and direct numerical simulation (DNS) of the compressible continuum model for the evolution starting from the Taylor-Green field (TGF[1]). For RSF, however, specific rules at the micro- or meso-scopic level need to be found and implemented in the computation, which, among other issues such as the (virtual) molecule number and noise of DSMC as addressed in Ref. 8 (see also the discussions of simulations of incompressible TGF evolution by discrete unified gas-kinetic scheme, pseudo-spectral and lattice Boltzmann methods[9]), again is not obvious. Both such types of algorithms are interesting and probably challenging, and, definitely deserve to be pursued. Here, as a first step, we discuss a semi-analytical algorithm, for DNS of RSF, based on the evolution equation dynamically projected from the full Navier-Stokes with the explicit analytical expressions for the precise relations.

For physical interest, RSF has been suggested to be possibily playing the role in general turbulence dynamics as special relativity in general relativity and, particularly, the helical RSF were proposed to be the 'chiral base flow (CBF)' for understanding the helicity effect on the compressibility of flows of neutral and ionized gases[10]. It is thus natural to numerically examine the very basic issues such as the production of small RSF eddies, a praradigm with TGF[1,11] extended also to different numerical concerns of compressible flows[12,13], and to test the relevant resutls in Ref. 10, such as the 'fastening' (compressibility-reduction) effect of helicity (compared to achiral base flow — aCBF). An even more detailed programme is to explore the (different) analytical structures of CBF and aCBF, such as the complex singularities[14], which should in some sense provide the mathematical foundation of the 'fastening' notion. Such a project however requires high-resolution and highly precise numerics, because, according to Ref. 14, the possible nonuniversality should present in the small-scale dissipation range. We indeed have relevant arguments and other high-resolution numerical data implying that the helicity weakens the overall complex singularities (by pushing them further away from the 'real world' and/or reducing the strength), numerical data of RSFs were however not available before. Thus, we first of all need an effective algorithm that can be used in conjunction with high-fidelity schemes and/or methods, which leads to the main purpose of this work to offer the proof of concept with working formulation, precise algorithm, solid examples and fundamental results.

Since the compressible RSF contains both 2D and 3D components (which can still be further decomposed into transversal and longitudinal modes) and is associated to rotation, as remarked earlier, it is intriguing what the genuine spectral transfer nature is. We know that when there is stratification, rotation or compressibility, the very genuine 2D spectral transfer dynamics in a simple ideal condition (with periodicity and incompressibility, say) may actually be facilitated with other ingredients in the much more complicated situation, leading to seemingly strange but actually natural phenomena[15–17]. It is precisely the wide connections of RSF that could make it be sensitive to a number of conditions, including the initial field and the pumping mechanism, among others, that setup the problem. The specific scenario then depends on the details. We thus should be very careful on what really is 'genuine' and need a lot of numerical experiments for examination.

The 'nature' (concerning spectral transfer, say) of triadic interactions of the dynamic equations (of quadratic nonlinearity, say) should be determined by *i)* the structure of the equation, *ii)* the structures of the excitations of the interacting modes, *iii)* the degrees ('energy'/$L^2$-norm levels, say) of excitations of the interacting modes. Waleffe[18] was able to clarify with the helical decompostion technique '*i*' and '*ii*' for Navier-Stokes and proposed the *instability assumption*, which, to our understanding, missed '*iii*' and is incomplete. For example, for 'free' relaxation without external disturbance, the initial distribution of excitations obviously should affect the 'nature' of mode interactions at that moment and later on, which can be taken into account in the statistical arguments for determining the 'fate' of final statistical equilibrium and, to some degree, the nonequilibrium transfer properties with careful considerations[19,20]. [Just as Kraichnan's own extension to the weakly compressible situation[21] of the incompressible absolute equilibrium analysis[22], one can straightforwardly perform the same calculation for 2D compressible flows with (approximate) energy invariance, but without the enstrophy invariance, and could naively conclude from the equipartition of energy that energy would transfer forwardly, which may not be completely wrong but is not fully in accord with the numerical details (e.g., Ref. 23 and references therein): there can be rich and significant (in terms of energy level, space and time scales etc.) nonequilibrium dynamical processes, such as the flux-loop[15] which is indeed forwardly 'leaking' due to dissipation at small scales, beyond the implication from simple absolute equilibrium analysis.] When there is external pumping, the acceleration keeps participating in modifying the structures (*ii*) and degrees (*iii*) of some of the interacting modes and then shapes the final (quasi-)steady state, in the case of no periodicity or recurrence. Thus, for the driven/accelerated case, from the long time point of view, it is the combination of *i)* the structure of the equation and *iv)* the nature of the pumping that determines the spectral transfer and final distribution properties. Ref. 3 exposed some mathematical structures of RSF, part of which will exhibit the power in the numerical computations presented here, but is still far from enough for complete theoretical understanding of the dynamics. Thus, further mathematical analysis, especially in the 'phase', say, Fouirier space, are much desired. Also, we have experimented with various acceleration schemes, focusing on the spatial-variation (uni- or multi-dimension) and helicity issues, including the additional considerations of randomness, but here we will demonstrate only part of the results closely related to our theme (more discussions on other relevant issues will be communicated elsewhere[24]).

Below, Eqs. (1a and 1b+), together with Eqs. (1b- and 1c), are established to compute isothermal RSF in a cyclic box with the semi-analytical algorithm. We will also find Eq. (1b) alternative to Eqs. (1b- and 1b+). Besides the decaying cases, also discussed are the driven RSFs with accelerations in the fashion of Mininni et al.[25], including a more detailed analysis of the spectra with respect to the particular scenario (intimate to that of flux-loop in 2D compressible turbulence[15] and to that speculated 2D vortical spectral transfer assisted by the wave modes[26]) of co-existing forward 'energy' and (2D) enstrophy transfers. Such a scenario supports our creation of the Jiu-Gong/Ba-Gua (ditetragonal/octagonal) pattern of cyclones resembling the recently photographed cluster encircling the northern polar of Jupiter.[27,28]

## II. FORMULATION AND ALGORITHM

### A. Starting from NSF

Let's start with the Euler equation in $\mathbb{E}^3$ for the RSF with (nondimensional) density $\rho$, pressure $p$, velocity $\boldsymbol{u}$,

$$\partial_t \rho = -\nabla \cdot (\rho \boldsymbol{u}) =: {}^0\!RHS, \tag{1}$$

$$\partial_t \boldsymbol{u}_h = -\boldsymbol{u}_h \cdot \nabla_h \boldsymbol{u}_h - \rho^{-1}\nabla_h p =: {}^h\!\boldsymbol{rhs}, \tag{1b-}$$

$$\partial_t u_3 = -(\boldsymbol{u}_h \cdot \nabla_h u_3 + u_3 u_{3,3}) - \rho^{-1} p_{,3} =: {}^3\!rhs, \tag{1c}$$

where $x_1$ and $x_2$ are the '$_h$orizontal' coordinates and the corresponding $\boldsymbol{u}_h := \{u_1, u_2\}$ is independent of the 'vertical' coordinate $x_3$, i.e.,

$$\boldsymbol{u}_{h,3} \equiv 0. \tag{2}$$

[The label (1b-) for an equation in the above is particularly arranged to be logically coherent with (1a, 1b+ and 1b), respectively in Sec. II B, II C and III B 2 below, for our governing equations, semi-analytical algorithm and numerical method of RSF.] For the barotropic case,

$$(\nabla p)/\rho = \nabla \Pi \tag{3}$$

where $\Pi$ is the specific enthalpy, and the isothermal (constant-temperature) relation $p = c^2 \rho$ results in

$$\nabla \Pi = c^2 \nabla \ln \rho, \tag{4}$$

where $c$ is the sound speed.

Internal viscosity (resp., external acceleration) models of $\boldsymbol{M}_h$ and $\mathscr{M}_3$ (resp., $\boldsymbol{a}_h$ and $a_3$) can be added to (1b- and 1c) respectively, and our objective is then to find the numerical solutions of Eqs. (1, 1b- and 1c) constrained by Eq. (2), which in general requires appropriately designed algorithms for performing consistent computations to obtain the accurate results. In this note, we are not interested in very complicated situations or sophisticated computations, but would rather look for reasonable simplications and semi-analytical treatments to compute RSF.

The viscosity model $\boldsymbol{M}(\boldsymbol{u})$ is determined by the constitutive relation of the medium. For the (generalized) Newtonian fluid, there is a dilational or 'second viscosity' coefficient whose neglection leads to the Stokesian model (thus the Navier-Stokes)

$$\boldsymbol{M}(\boldsymbol{u}) = \mu\rho^{-1}\nabla^2\boldsymbol{u} + \frac{\mu}{3\rho}\nabla(\nabla \cdot \boldsymbol{u}) \tag{5}$$

when the dynamical viscosity $\mu$ is constant. For simplicity, we will focus on the model $\boldsymbol{M}(\boldsymbol{u}) = \nu\nabla^2\boldsymbol{u}$ applied in incompressible flows with constant kinetic viscosity $\nu$. This is not always very realistic, but can be a good approximation in many physical situations, unless the flow is extremely compressible with very high tempearture, say. In such a case,

$$\boldsymbol{M}_h = \nu\nabla_h^2\boldsymbol{u}_h, \text{ and } M_3 = \nu\nabla^2 u_3. \tag{6}$$

Note that for analytical[3,21] and/or numerical (e.g., Refs. 29 and 30) convenience, Eq. (1) has also been written for the 'logarithmic variable', $\ln\rho$,

$$\partial_t \ln\rho = -\boldsymbol{u} \cdot \nabla \ln\rho - \nabla \cdot \boldsymbol{u} =: {}^0\!rhs, \tag{7}$$

which is equivalent except for the vacuum solution that we are not interested in here.

On the other hand, particularly for the computational fluid dynamics (CFD) formulation and programming, and, for better behavior of shock-capturing schemes (e.g., Ref. 31), it is also useful to write in the following *conservative form* for the Navier–Stokes equations of the compressible flow of an ideal gas

$$\begin{cases} \boldsymbol{U}_{,t} = -\boldsymbol{F}_{j,j} + \boldsymbol{V}_{j,j} =: \boldsymbol{RHS}, \\ p = \rho\mathcal{R}T, \end{cases} \tag{8}$$

where

$$\boldsymbol{U} = \begin{pmatrix} \rho \\ {}^h\boldsymbol{U} \\ \rho u_3 \\ E \end{pmatrix}, \quad \boldsymbol{F}_j = \begin{pmatrix} \rho u_j \\ \rho u_1 u_j + p\delta_{1j} \\ \rho u_2 u_j + p\delta_{2j} \\ \rho u_3 u_j + p\delta_{3j} \\ (E+p)u_j \end{pmatrix}, \quad \boldsymbol{V}_j = \begin{pmatrix} 0 \\ \sigma_{1j} \\ \sigma_{2j} \\ \sigma_{3j} \\ \sigma_{jk}u_k + \kappa T_{,j} \end{pmatrix}, \tag{9}$$

with ${}^0\boldsymbol{U} := \rho$ and ${}^h\boldsymbol{U} := \begin{pmatrix} \rho u_1 \\ \rho u_2 \end{pmatrix}$ to which correspond the ${}^0\boldsymbol{RHS}$ and ${}^h\boldsymbol{RHS}$ (for later reference), respectively, in $\boldsymbol{RHS}$. Here, $\delta_{ij}$ is the Kronecker delta, and the summation over repeated indices has been assumed; $\kappa$ is the conductivity of the temperature $T$, and $\mathcal{R}$ is the ideal gas constant. The total energy is $E = \frac{p}{\gamma-1} + \frac{1}{2}\rho u_j u_j$ where $\gamma \equiv C_p/C_v$ is the adiabatic exponent, the ratio of the constant-pressure heat capacity $C_p$ to the constant-volume one, $C_v$. The viscous stress $\sigma_{ij} = \mu(u_{i,j} + u_{j,i}) - \frac{2}{3}\mu\theta\delta_{ij}$, in general also accounts for the dilatation $\theta := \nabla \cdot \boldsymbol{u}$. The dynamic viscosity $\mu$ assumes Sutherland's law[31] $\mu = \frac{1.4042T^{1.5}}{T+0.40417}\mu_\infty$ with $\mu_\infty = 1.716 \times 10^{-5}$kg/(m·s): $\mu$ is indeed constant when $T$ is not varying (isothermal), which is a good approximation for moderate cases free of boundary.

## B. The governing equations for RSF

The RSF has not only the defining characteristics in the velocity but also some particular thermodynamic structures which govern the dynamics. Below, we will derive the results from the 2C2Dcw1C3D velocity field for the 3-space dynamics (more complete analytical results, including those for high-dimensional space and Lie invariances, can be found in Ref. 3).

Taking derivative with respect to $x_3$ in Eq. (1b-), Eq. (2) indicates[32]

$$[(\nabla_h p)/\rho]_{,3} = 0 \tag{10}$$

which, together with Eq. (3), turns the barotropic Eq. (10) into

$$(\nabla_h \Pi)_{,3} = 0 \text{ (or } \Pi_{,13} = \Pi_{,23} = 0). \tag{11}$$

In words, $\Pi$ should be decomposed into two functions, $\mathscr{P}_h$ and $\mathscr{P}_3$, one of only the horizontal coordinate $\boldsymbol{x}_h := \{x_1, x_2\}$ and the other of only $x_3$:

$$\Pi = \mathscr{P}_3(x_3) + \mathscr{P}_h(x_1, x_2). \tag{12}$$

We now consider the RSF in a box of dimension $L_z \times L_2 \times L_3$, cyclic in each direction, or with $L \to \infty$ in some direction(s) with the field vanishing sufficiently fast. Introducing

$$\langle \bullet \rangle_{12} := \frac{\int_0^{L_2}\int_0^{L_1} \bullet \, dx_1 dx_2}{L_1 L_2}, \quad \langle \bullet \rangle_3 := \frac{\int_0^{L_3} \bullet \, dx_3}{L_3} \text{ and } \langle \bullet \rangle_{123} := \frac{\int_0^{L_1}\int_0^{L_2}\int_0^{L_3} \bullet \, dx_1 dx_2 dx_3}{L_1 L_2 L_3}, \tag{13}$$

we have

$$\langle \Pi \rangle_3 = \mathscr{P}_h(x_1, x_2) + \langle \mathscr{P}_3 \rangle_3, \tag{14a}$$

$$\langle \Pi \rangle_{12} = \mathscr{P}_3(x_3) + \langle \mathscr{P}_h \rangle_{12}, \tag{14b}$$

$$\langle \Pi \rangle_{123} = \langle \mathscr{P}_h \rangle_{12} + \langle \mathscr{P}_3 \rangle_3 = \langle \Pi \rangle_{12} + \langle \Pi \rangle_3 - \Pi,$$

$$\text{i.e., } \Pi = \langle \Pi \rangle_{12} + \langle \Pi \rangle_3 - \langle \Pi \rangle_{123}. \tag{14c}$$

We take, with Eq. (4) for the isothermal case, the specific enthalpy particularly (ignoring the irrelevant constant)

$$\Pi = c^2 \ln \rho = c^2(\langle \ln \rho \rangle_{12} + \langle \ln \rho \rangle_3 - \langle \ln \rho \rangle_{123}). \tag{15}$$

We then bring Eqs. (15 and 14c) into Eq. (7), with the interchangeability of the order of the time derivative and the average operators defined in Eq. (13) and with $\langle u_{1,1} \rangle_1 = \langle u_{2,2} \rangle_2 = \langle u_{3,3} \rangle_3 = 0$ from the periodic boundary condition, to obtain the (partial) integral-differential equation

$$\partial_t \ln \rho = \langle ^0 rhs \rangle_{12} + \langle ^0 rhs \rangle_3 - \langle ^0 rhs \rangle_{123} \tag{1a}$$

Obviously, the above derivation works also for the case with a 2C2Dcw1C3D external acceleration $\boldsymbol{a}$ exerted on Eqs. (1b- and/or 1c), with the horizontal component $\boldsymbol{a}_h$ being independent of $x_3$. Eq. (1a) applies in more general nonbarotropic RSFs, but we focus only on the isothermal case in this note for the (numerical) proof of concept.

## C. The strategy and algorithm for simulations

Our algorithm for integrating our partial-integral-differential equations can be developed from the standard method of CFD. The 'typical' strategy adopted here is based on the observation from Eqs. (1a, 1b- and 1c) that the 2C2Dcw1C3D RSF can be regarded as system consisting all original NSF components with additional difining and derived structures of RSF, thus a two-step procedure, with *the first (standard CFD) part* in the conventional NSF solver and *the second (RSF-specialization) part* with the additional RSF structure implemented when needed, is adopted. An alternative 'customized' strategy would be to treat RSF as a completely new system, writting out the presenting components and compute them accordingly, which is not exercised here but will be further remarked in Sec. V B.

As can be seen from the governing equations, particularly (1a), of RSF dynamics established in the above, *the first part* involves the familiar discretization and computation of the compressible Navier-Stokes quation, in which, as mentioned, the wisdom is that the conservative Eq. (8) in which, for instance, Eq. (1) instead of (1a), is appropriate for use in the case of high speed flows with shocks needed to be captured accurately[31]: Eqs. (7 and 1a) are not in the conservative form. The latter can be directly used for discretizing and integrating weakly compressible flows[30] but in general do not quite satisfy our interests of the flows also involving (strong) shocks. So, we need a combination of the two methodologies, both appear to be most conveniently realized by the mature high-order finite difference schemes, such as the compact and weighted essentially non-oscillator (WENO) schemes[33] used in various relevant studies[34–37].

Our strategy is to compute Eqs. (7 and 1a) in an actually-conservative way, in the sense that the $^0rhs$ of Eq. (7) is computed from the $^0RHS$ of Eq. (1), simply by $(\ln \rho)_{,t} = \rho_{,t}/\rho$. $\boldsymbol{u}$ can also be similarly treated (see below). Since the transformation (dividing by $\rho$) involves only the local variable without derivative and $\rho$ is nonvanishing (we won't study the problem with vacuum in this note, as said), neither accuracy loss nor singular issue will arise.

With such a strategy, *the second part* then can be performed, after the (actually-conservative) fluxes are computed, with the spatial averaging operators defined in Eq. (13). Higher-order time accuracy, such as the Runge-Kutta methods, then can be applied for (explicit) time marching, completing the implemention of a semi-analytical algorithm.

We remark that in principle Eqs. (1a) in Sec. II B, (1b- and 1c) in Sec. II A completely define (with initial 2C2Dcw1C3D field) the RSF without the necessity of imposing other precise analytical relations derived in the last section. However, due to numerical errors in computing $\boldsymbol{u}_h$ with Eq. (1b-) or obtaining $\boldsymbol{u}_h$ through $\boldsymbol{u}_h = \rho\boldsymbol{u}_h/\rho$ from the computation of the conservative Eq. (8), i.e., the second and third components of (8 and 9), it is practically needed to impose, say, Eq. (2) which reads in terms of computation

$$\boldsymbol{u}_h = \langle \boldsymbol{u}_h \rangle_3. \tag{1b+}$$

We will come back to this point in Sec. III B.

### III. THE PHYSICAL PROBLEM AND NUMERICAL METHODS

A most fundamental problem in fluid turbulence is the development of multi-scale structures and small-scale dissipation, associated to the singularities (complex or real). We focus on the isothermal RSFs in a cyclic box of dimension $2\pi \times 2\pi \times 2\pi$, without loss of generality: even for boxes of non-unit aspect ratio(s), normalization to such a box is still appropriate (with however other coefficients, such as the viscosity for different velocity components, being accordingly changed). A paradigm has been studying the evolution of the so-called

Taylor-Green field (TGF)[1] and Orszag-Tang field (OTF)[38]. Indeed, much further has been carried forward along this way for incompressible (e.g., Refs. 11 and 39, respectively for TGF and OTF) and also compressible flows, for the latter of which it has even been proposed as the 'benchmark' case for numerical methods (e.g., Ref. 12 and 40, respectively for TGF and OTF). The OTF studies have been also carried out in 3D magnetohydrodynamics (MHD) (e.g., Ref. 41). Though our studies may be extended to plasma flows[10], we shall restrict ourselves in the neutral gas, evolving from the initial prototypical fields in Sec. III A (also used as the accelerations in the driven case in Sec. V A) with the numerical method detailed in Sec. III B. And, corresponding to OTF, actually, as we shall see, it appears more precise to consider the Arnold-Beltrami-Childress flow (ABCF[42]).

We will freely apply the correpondences $x \leftrightarrow x_1$, $y \leftrightarrow x_2$ and $z \leftrightarrow x_3$ for historical reason associated to TGF and ABCF, and $\boldsymbol{\omega} := \nabla \times \boldsymbol{u}$ denotes the vorticity of the velocity $\boldsymbol{u}$.

## A. Prototypical flows

The TGF (after appropriate normalizations and reparameterization with a rotation angle $\theta$[11]) and ABCF read, respectively,

$$^{TGF}\boldsymbol{u} = \begin{pmatrix} \frac{\sin(\theta+2\pi/3)}{\sqrt{3}} \sin x \cos y \cos z \\ \frac{\sin(\theta-2\pi/3)}{\sqrt{3}} \cos x \sin y \cos z \\ \frac{\sin\theta}{\sqrt{3}} \cos x \cos y \sin z \end{pmatrix}, \quad ^{ABCF}\boldsymbol{u} = \begin{pmatrix} A \sin z + B \cos y \\ C \sin x + A \cos z \\ B \sin y + C \cos x \end{pmatrix}, \quad (16)$$

where OTF (with an extra $u_z$ component) corresponds to $A = 0$, $B = 1$ (with $y$ replaced by $y + \pi/2$) and $C = 1$.

Before proceeding, it is better putting some terminologies in to the following

**Definition 1** *A vector field $\boldsymbol{v}$ is 'helical' if the helicity $\int\int\int \nabla \times \boldsymbol{v} \cdot \boldsymbol{v} d^3\boldsymbol{x} \neq 0$ and is 'purely helical' ('unichiral') if each of the Fourier component $\hat{\boldsymbol{v}}(\boldsymbol{k}) = \int\int\int \boldsymbol{v}\exp\{-\hat{i}\boldsymbol{x}\cdot\boldsymbol{v}\} d^3\boldsymbol{x}$ is a 'helical mode', the latter meaning that $\hat{i}\boldsymbol{k}\times\hat{\boldsymbol{v}} = \pm k\hat{\boldsymbol{v}}$ with $\hat{i}^2 = -1$. The $+$ and $-$ signs are usually assigned respectively to right- and left-handnesses. $\boldsymbol{v}$ is 'homochiral', if the $+$ or $-$ sign applies uniformly for every $\boldsymbol{k}$, and 'Beltramian' if $\exists \kappa$, $\kappa^2 > 0$, $\nabla \times \boldsymbol{v} = \kappa \boldsymbol{v}$.*

The local helicity density of TGF vanishes everywhere, $^{TGF}\boldsymbol{\omega} \cdot {}^{TGF}\boldsymbol{u} = 0$, while ABCF is *purely helical*, actually *Beltramian*[42]. The perturbative calculations similar to that of Taylor and Green[1] appears formidable, if possible to some degree with our RSF being possibly more analytically tractable, for the compressible case, because the density modes enter and the pressure can not simply be determined by the Poisson equation, even if initially is.

### 1. *2C2Dcw1C3D*

We first consider the TGF fashion RSF velocity

$$\boldsymbol{u} = \begin{pmatrix} \cos x \sin y \\ -\sin x \cos y \\ \sin x \sin y \cos z \end{pmatrix}. \qquad (17)$$

It appears in Fig. 2 for the field pattern that the isosurfaces of small-amplitude vorticity are highly three dimensional while those of high amplitude are only weakly depending on $z$.

$\boldsymbol{u}$ is already fully 2C2Dcw1C3D, thus the simplest initial density can just be uniform $\rho(t_0) = 1$. We can of course also compute from the isothermal relation the one corresponding to which the pressure gradient balances the 'parallel' (to wavevector) component of the nonlinear advection term, as is usually done in TGF-relevant studies: such balance however, unlike in the incompressible flow, is not prereserved, thus not necessary for the purpose of this note. Different initial pressure/density fields relax and decay differently, we however shall not digress into such studies.

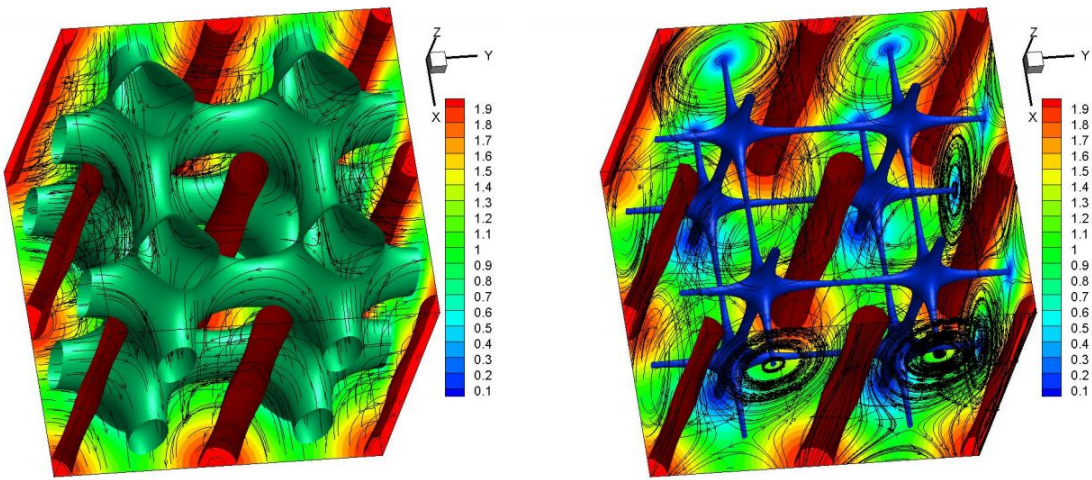

FIG. 2. The vorticity strength, $|\boldsymbol{\omega}|$, slices and isosurfaces (0.7 and 1.9 in the left panel, and, 0.1 and 1.9 in the right one), and, velocity (left) and vorticity (right) streamtraces.

Nevertheless, problems such as the flow compressibility difference between CBF and aCBF raise the issue of 'fair comparison': when CBF and aCBF have the same transversal and parallel (to wavevector) powers, and, the same density field, their initial acceleration fields are then different. Thus, absolutely 'fair' comparison is in general impossible, and what can be pursued is to be essentially fair in exposing the physical issue in concern. We will come back to this point in Sec. III A 2 for the fields related to the following helical decomposition.

The local helicity density $\boldsymbol{\omega} \cdot \boldsymbol{u} = 0$ for $\boldsymbol{u}$ in Eq. (17). We then can compute the purely helical, say, the right-handed $^R\boldsymbol{u}$ from the standard helical decomposition of the Fourier coefficients[43,44] (explicit formula given in Sec. III A 2), but, unfortunately, such $^R\boldsymbol{u}$ is not an RSF. It is of course easy to design helical but not purely helical RSFs, as for the acceleration to be used. For purely helical RSF, we have to turn to the field independent of the vertical coordinate $z$, that is a three-component-two-dimensional (3C2D) velocity, due to the following

**Theorem 1** *A 2C2Dcw1C3D but not 3C2D flow can not be 'purely helical'.*

*Proof.* $\boldsymbol{u}_{h,3} = 0$ means $\hat{u}_1(\boldsymbol{k}) = \hat{u}_2(\boldsymbol{k}) = 0 \ \forall \ k_3 \neq 0$. The purely helical, say, without loss of generality, the right-handed Fourier mode $\hat{\boldsymbol{u}}_h$, satisfies however, e.g., $\hat{i}(k_1 \hat{u}_2 - k_2 \hat{u}_1) = k \hat{u}_3$ by definition 1, and contradiction results if $k\hat{u}_3(\boldsymbol{k}) \neq 0$ for $k_3 \neq 0$. □

The ABCF in Eq. (16) with $A = 0$ but $BC \neq 0$ is a Beltramian 3C2D RSF.

*2. 3C2D*

To have a comparison between purely neutral (with right- and left-handed sectors exactly balanced) and purely helical RSF initial fields, we then consider

$$\boldsymbol{u} = \begin{pmatrix} \cos x \sin y \\ -\sin x \cos y \\ \sin x \sin y \end{pmatrix}, \ ^R\boldsymbol{u} = \begin{pmatrix} \frac{1}{2}\left(\sqrt{2}\cos x \sin y + \sin x \cos y\right) \\ \frac{1}{2}\left(-\sqrt{2}\sin x \cos y - \cos x \sin y\right) \\ \frac{1}{2}\left(\sqrt{2}\sin x \sin y - 2\cos x \cos y\right) \end{pmatrix}. \quad (18)$$

In the above, $^R\boldsymbol{u}$ is computed from the helical decomposition of Fourier coefficients[43,44],

$$^R\hat{\boldsymbol{u}} = (\hat{\boldsymbol{u}} + \hat{i}\boldsymbol{k} \times \hat{\boldsymbol{u}}/k)/\sqrt{2}, \ ^L\hat{\boldsymbol{u}} = (\hat{\boldsymbol{u}} - \hat{i}\boldsymbol{k} \times \hat{\boldsymbol{u}}/k)/\sqrt{2} \quad (19)$$

of the incompressible $\boldsymbol{u}$, representing the purely right-handed helical sector. The reason we do this is because, just like TGF, here $(\nabla \times \boldsymbol{u}) \cdot \boldsymbol{u} = 0$. We will compare the evolutions from such fields, both augmented with, say,

$$\rho(t_0) = \rho_0(1 + \epsilon \cos z) \tag{20}$$

where $\rho_0$ (= 1 here) is the reference density and $1 > |\epsilon| \neq 0$. The pressure $p = c^2 \rho$ then will drive the isothermal flow to be 2C2Dcw1C3D during which small eddies are produced.

Both velocities are incompressible and $\int_0^{2\pi} \int_0^{2\pi} \boldsymbol{u}^2 \, dxdy = \int_0^{2\pi} \int_0^{2\pi} {}^R\boldsymbol{u}^2 \, dxdy$, for which, in Eq. (19), we have particularly replaced the factor $1/2$ in the stardard formular[43,44] with $1/\sqrt{2}$. However, the two chiral sectors of vorticities of $\boldsymbol{u}$ can cancel, leading to much smaller minimum ($\approx 0$) of $|\nabla \times \boldsymbol{u}|$ than that ($\approx 1.4$) of ${}^R\boldsymbol{u}$, while the maxima ($\approx 2.0$) of $\boldsymbol{u}$ is larger (due to mutual enhancement) than that ($\approx 1.0$, in Fig. 3) of ${}^R\boldsymbol{u}$, which already indicates the insufficiency or incompleteness of the traditional terminology in compressible or aeroacoustic turbulence context[21] of 'vortical' or 'shear' mode (compared to the dilational/compressive mode, c.f., Ref. 48 and references therein). Fig. 3 compares the vorticity strength and velocity streamlines of the two initial fields, showing the different flow patterns, though the vorticity strength patterns look similar (but distinct quantitatively). Also presented in the right panel is the computed helicity density and the vorticity-streamline pattern: it looks the same as the middle panel, because ${}^R\boldsymbol{u}$ is a Baltrami field due to the fact that the two modes of $\boldsymbol{u}$ has the same wavelength $k = \sqrt{2}$. For more details, this Beltrami field satisfies $\nabla \times \boldsymbol{u} = k\boldsymbol{u}$, and actually, corresponding to the maximum and minimum figures of vorticity amplitudes given in the above, the maximum of helicty is indeed $1.41^2/\sqrt{2} \approx 1.41$ while the minimum, again indeed, $(1.0)^2/\sqrt{2} \approx 0.71$ [the maxima and minima given the legends for contours are in general close to but not exactly the actual ones]. In other words, we already have the OTF- or ABC-fashion initial field given in the second half of Eq. (16) to which only a variable tranformation is needed from the latter of Eq. (18).

The above set up of CBF and aCBF appear to be good for comparing the evolution of compressibility. The initial acceleration fields and the changing rates of velocity divergence are in general different, from the beginning to the end, resulting in the physical conclusions in Sec. IV B consistent with the 'fastening' notion[10]. We believe it is appropriate to conclude the different consequences on flow compressibility from such essentially fair physical conditions. Similarly is the situation with random initial fields and/or forcing, in which case it is obvious impossible to have all details to be the same, except for the relevant statistics, for a comparison of turbulence. Also, random or turbulent fields with and without helicity do not necessarily present the obvious large-scale or global-pattern difference as in Fig. 3 for the simple flow supperposed by a few Fourier modes, because plenty of intermediate- or large-$k$ modes can contribute a lot to the helicity. Detailed discussions of the latter however belong to another communication.

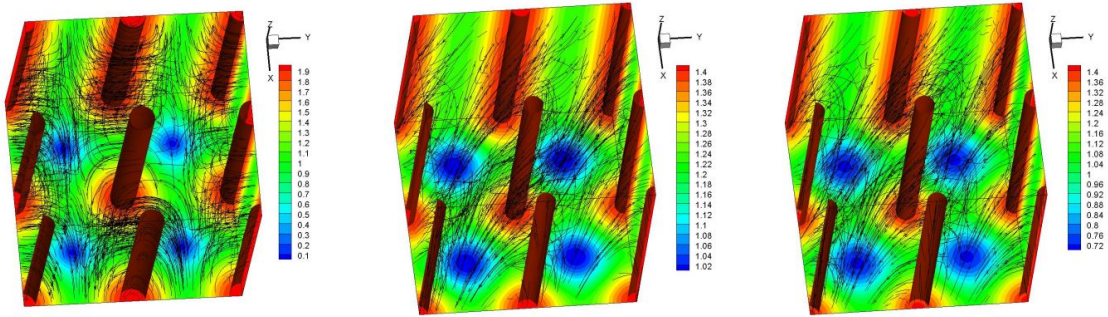

FIG. 3. The vorticity strength ($|\nabla \times \boldsymbol{u}|$) slices and isosurfaces and, velocity streamlines of $\boldsymbol{u}$ (left) and $^R\boldsymbol{u}$ (middle), the slices, isosurfaces, helicity density and vorticity streamlines of the latter is also presented in the right pannel (exactly the same/isomorphic to the middle pannel due to the Beltramity).

## B. Numerical method

As remarked in the end of Sec. II C, the problems in Sec. III A with initial data of RSF should be solved with the equations established there. However, though the problem and mathematical description are already complete, the standard CFD procedure and the RSF-specialization, such as the extra imposition of Eq. (2), actually contain important technical details in the numerical method and need further clarification.

### 1. The standard CFD part

In the numerical implementation, the first part, as mentioned in Sec. II C, for our purpose can in principle any suitable standard CFD method, in the prodecure of which Eq. (8) is normalized by the characteristic length $\ell^*$, velocity $u^*$, time $L/U$, density $\rho^*$, temperature $T^*$, pressure $\rho^* \mathcal{R} T^*$, kinetic energy $\rho^* U^2$, sound speed $c^* = \sqrt{\gamma \mathcal{R} T^*}$, dynamic viscosity $\mu^*$, and heat conductivity coefficient $\kappa^*$. The normalization results in non-dimensional parameters, including the Reynolds number $Re := \rho^* UL/\mu^*$, Mach number $Ma := U/c^*$, Prandtl number $Pr := \mu^* C_p/\kappa^*$, and adiabatic exponent of gas $\gamma$, and total energy now reads

$$E = \frac{p}{(\gamma-1)\gamma Ma^2} + \frac{1}{2}\rho u_j u_j. \tag{21}$$

Our isothermal case simply corresponds to that with unit value of $T$. We computed the nondimensionalized equation with various initial parameters with, for instance, the Mach number $Ma$ ranging from 0.1 to 2.0 for various tests and checks, but here only the simulations in a cube of dimension $2\pi$ resolved with $128^3$ uniform grids for $\gamma = 1.4$, $Pr = 0.7$, $Re = 450$ and $Ma = 1$ are reported, as the proof of concept ($Ma = 0.1$ case with deterministic accelerations will also be presented in Sec. V for futher discussions, while random RSFs, including those of $256^3$ and $512^3$ resolutions showing reasonable grid convergence at $128^3$, will be communicated elsewhere[24]). More detailed and specific studies of (complex) singularities, fully developed turbulence with random structures, among others which may require higher resolution and precision, and, even other approaches such as the pseudo-spectral method[45], deserve further studies in the future.

In the RSF solver for these specific computations, assembling of the conservation and nonconservation forms of the Navier–Stokes equation has been made. In particular, the equation for the logarithmic of density for time marching with its right-hand side however is computed from the method using the conservative form in terms of density itself. Using the logarithmic of density is a must for RSF as already indicated earlier and will be further explained below, while using the conservative form for discretization is to capture the shocks well[31]. Specifically for the finite-difference schemes, the viscous terms are discretized by the compact eighth-order finite difference scheme and the convective terms are discretized by the seventh-order weighted essentially non-oscillatory (WENO) schemes[36,37]. The time integration is advanced by the third–order total variation diminishing Runge-–Kutta method[46].

### 2. *The second RSF-specialization part*

Some more important details of the second RSF-specialization part follow: with

$$^h\boldsymbol{RHS} = (\rho \boldsymbol{u}_h)_{,t} = \rho \boldsymbol{u}_{h,t} + \rho_t \boldsymbol{u}_h = \rho\ ^h\boldsymbol{rhs}\ +\ ^0\!rhs\ \boldsymbol{u}_h \tag{22}$$

and $^0rhs = {}^0RHS/\rho$ defined in Eq. (7), we have

$$\boldsymbol{u}_{h,t} = {}^h\boldsymbol{rhs} = \left[{}^h\boldsymbol{RHS} - \boldsymbol{u}_h \, {}^0RHS/\rho\right]/\rho. \tag{23}$$

Given Eqs. (1a) in Sec. II B and (1b-) in Sec. II A and the initial RSF data, the RSF evolution is mathematically assured, however, with $^h\boldsymbol{RHS}$, $^0RHS$ and $\rho$ containing the $z$-dependence not numerically cleanly removed as in Eq. (1a), the computation of $^h\boldsymbol{rhs}$ in Eq. (23) needs clearing up the $z$-dependent errors. Thus, according to Eq. (1b+) in Sec. II C and the interchangeability of the order of time derivative and spatial averaging, we should instead compute

$$\boldsymbol{u}_{h,t} = \langle (\rho \, {}^h\boldsymbol{RHS} - {}^0RHS \, \boldsymbol{u}_h)/\rho^2 \rangle_3. \tag{1b}$$

That is, the 'numerically complete' governing equations are (1a, 1b and 1c), and solving such a system with whatever numerical discretization scheme may be considered as the 'semi-analytical' method, in the sense that the $\boldsymbol{u}_{h,3} \equiv 0$ property is automatically satisfied by the self-consistent dynamical equations and the errors from whatever numerical discretizations and integrations in this property is cleanly removed, up to the computer roundoff. Since it is purely the purpose of removing the numerical errors to perform (*i*) the computation of Eq. (1b), i.e., additional spatial averages over Eq. (23) and (*ii*) the imposition of Eq. (1b+) after computing $\boldsymbol{u}_h$ from $^h\boldsymbol{U}$ in Eq. (8) mentioned in the end of Sec. II C, there should be no difference between the latter two approaches, as indeed verified by numerical tests. For the current purpose of the problems set up in Sec. III A, the computation of (*ii*) is actually more economy, with less operations of spatial averages or Fourier mode truncations.

Simply imposing the extra Eq. (1b+) in each step of solving the original, say, the conservative-form Eq. (8), in the standard CFD may not be considered as completely 'wrong' but should be regarded as only a 'primitive' or 'zeroth order' scheme (which is possible to be improved with other techniques such as the dual-time stepping method[7]). Such a treatment does not take the self-consistent RSF dynamics with the accompanying precise structures in Sec. II B (see more in Ref. 3) into account. Since the response time to the operation (1b+) is finite, characterized in general by the sound speed $c$, it is hard to estimate the errors. Numerical experiments were performed with such 'zeroth order' algorithm against the 'semi-analytical' one, showing that the former leads to inaccurate results with uncontrollable growth of errors from the inconsistent response between the density/pressure and the imposed Eq. (1b+): Fig. 4 constrasts the typical structures of $\partial_x(\ln\rho)$ and $\partial^2_{xz}(\ln\rho)$ resulting from the two algorithms at some early stage of the simulations, showing that the 'zeroth order' algorithm is poorly in accord with Eq. (10) and that the 'semi-analytical' one is precise (up to the pure 'noise' of numerical errors from discretization and computer round off): the structures of $\partial_x(\ln\rho)$ from the former present obviously visiably deviatiations from perfect vertical 'bars' as those from the latter, and the difference of the errors in this respect is of 15 orders of magnitude, as indicated by the legends. Comparisons of $\partial_y(\ln\rho)$ and $\partial^2_{yz}(\ln\rho)$ are of similar character and not shown.

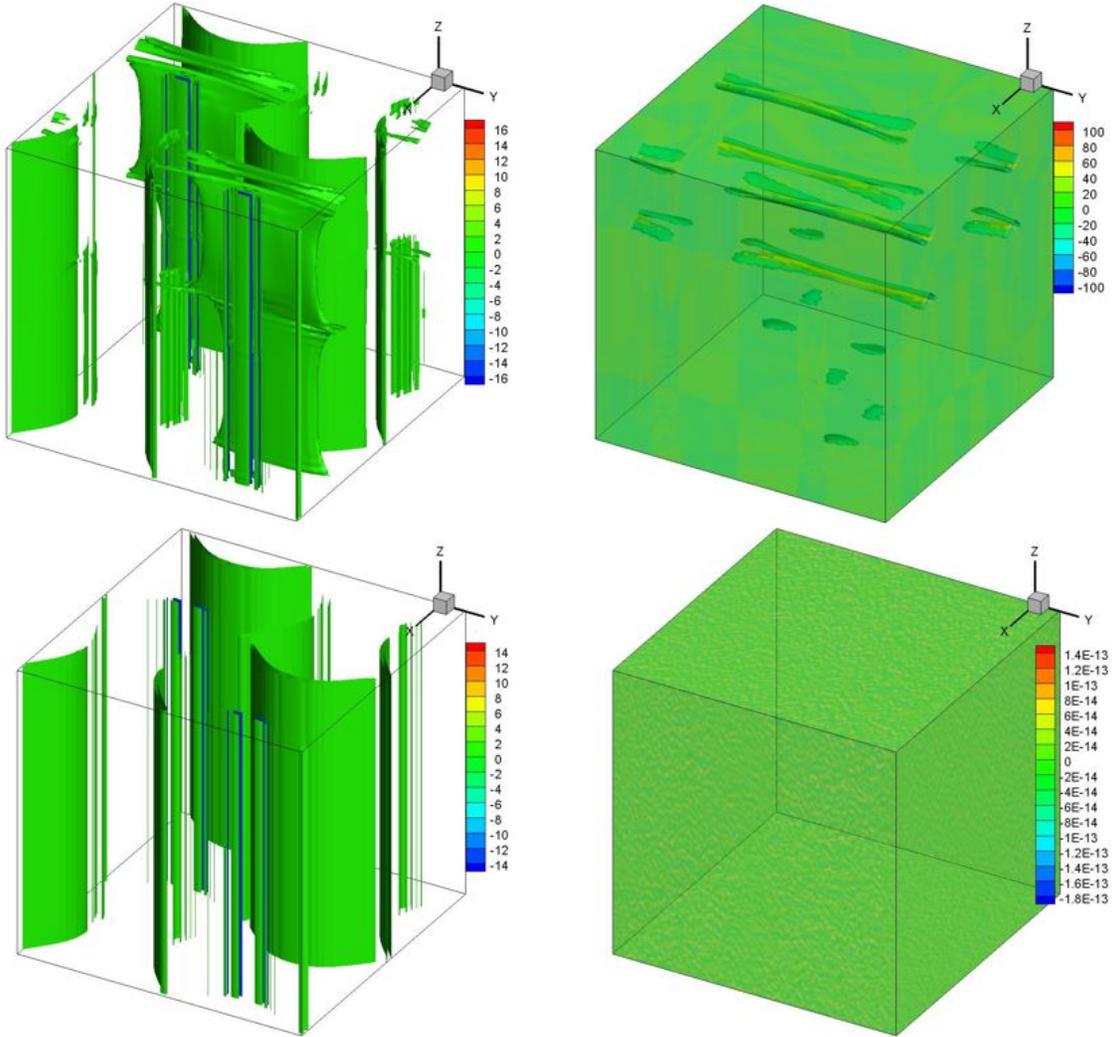

FIG. 4. Snapshots of the isosurfaces and contours (translucent) of $\partial_x(\ln\rho)$ (left) and $\partial^2_{xz}(\ln\rho)$ (right) for the 'zeroth order' algorithm (upper) and the 'semi-analytical' algorithm (lower) from test simulations of inviscid RSF starting from Eq. (17).

## IV. RESULTS

In this section we present the solutions to the physical problems formulated in Sec. III A. A comparion between RSF and the Navier-Stokes flow (NSF) is also made.

All simulations were performed with initial $Re = 350$ and $Ma = 1$.

### A. Multi-scale excitations of RSF eddies from 2C2Dcw1C3D initial fields

First of all, since the flow is completely new in the sense of realizations (*in silico*), we present in Fig. 5 for general conception the comparison of the fields at the very early time $t = 0.3$ between RSF and NSF starting from the same field of Eq. (17) and Fig. 1 at $t = 0$, showing very different evolution routes: obviously, NSF immediately generates 3D $\boldsymbol{u}_h$, together with other differences to RSF.

To see the differences when the systems maturize with 'fully developed' multi-scale excitations, we also present in Fig. 6 the NSF fields and spectra at $t = 5$ (left and middle panels) for comparison:

$$E(k) := \sum_{|\boldsymbol{k}|=k} \frac{|\hat{\boldsymbol{u}}(\boldsymbol{k})|^2}{2}, \quad {}^vE_v(k_v) := \sum_{|k_3|=k_v}^{\boldsymbol{k}_h} \frac{|\hat{u}_3(\boldsymbol{k})|^2}{2} \text{ and } {}^hE_h(k_h) := \sum_{|\boldsymbol{k}_h|=k_h}^{k_3} \frac{|\hat{\boldsymbol{u}}_h(\boldsymbol{k})|^2}{2} \quad (24)$$

where $\boldsymbol{k}_h := \{k_1, k_2\}$ which can be 'trivially' extended to be a 3-space vector $\{k_1, k_2, 0\}$. Actually, we notice that at such a moment the system has not really reached the 'mature' state, with an approximate $k^{-5/3}$ (which is generally referred to the Kolmogorov 1941 – K41 – law[47]) scaling in the potentially inertial range, which happens at $t = 6$ as shown by the right panel: It appears that all NSF spectra maturize (approximately) to the $k^{-5/3}$ law in the potentially inertial range where large-scale forcing (if exists) and small-scale damping effects are negligible. Although multi-scale excitations present and the spectra appear like those of the conventional turbulence, we note that obvious deterministic order represented by the 'precise' (up to numerical errors) symmetry of the pattern inherited and developed from the original field is in the flow. Such a characteristic, shared by all the results presented in this note, is not that of 'statistical symmetry' in the conventional (multi-)fractal or fully developed turbulence theory[47]. This observation also means that it makes sense to compare the definite flow structures of our simulations at the same moment. We will nevertheless make remarks related to 'turbulence' during the discussions of our results, and we believe

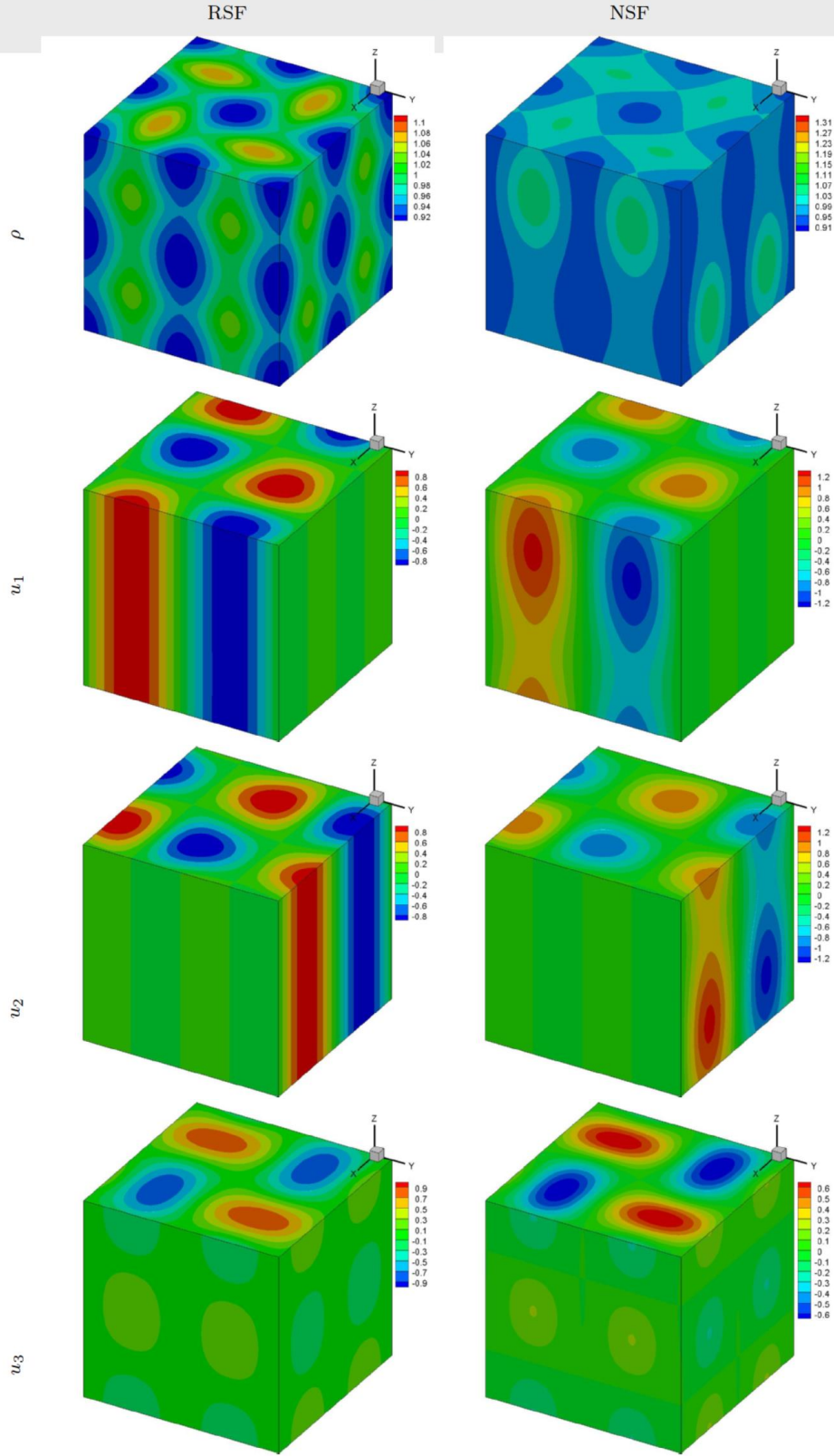

FIG. 5. Contours of the primitive variables $\rho$, $u_1$, $u_2$ and $u_3$ at $t = 0.3$, for NSF and RSF.

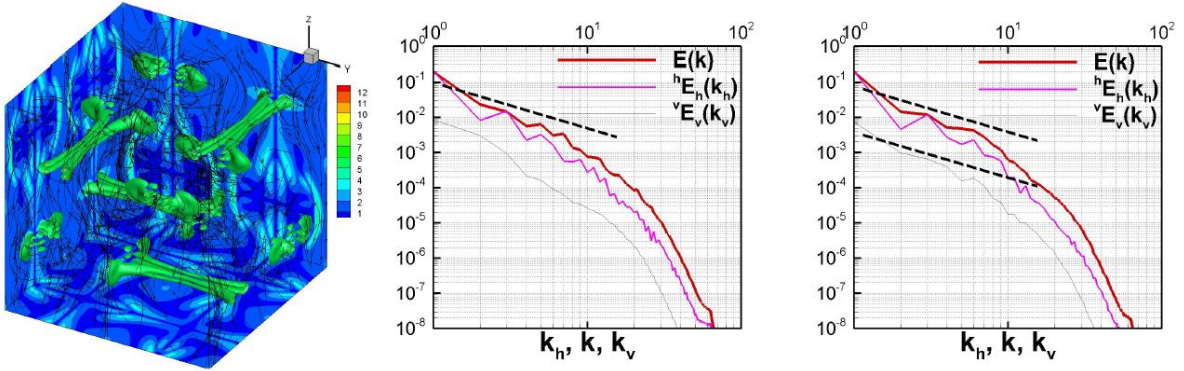

FIG. 6. Isosurfaces, slices of vorticity amplitude and streamtraces of vorticity (left), and, the power spectra of velocity (middle) of NSF at $t = 5$. Also presented are the corresponding spectra at $t = 6$ (right): modes of noninteger $k$ are always grouped to the nearest integer shell, thus those of $k = \sqrt{2}$ are counted on the first shell, if not particularly pointed out as in Fig. 8 and Sec. V. Dashed lines denoting $k^{-5/3}$ law are for reference.

they indeed are relevant.

Very differently, as shown in Fig. 7 (the power spectra and fields correspond in the two upper rows for three different simulations for the same flow, respectively, to those of Fig. 6), RSF runs into a completely different state, both in terms of patterns and velocity power spectra. The three columns are respectively from three different simulations with band-optimized symmetric WENO[37] (WENO-SYMBO: left) and WENO-Z[36] (middle) for the RSF with Stokesian viscosity without the compressibility effect, and, with WENO-SYMBO for the RSF with full Stokesian viscosity. These results indicate on the one hand that the differences from the details of the different numerical schemes are small and irrelevant to our discussions, and, on the other hand, our semi-analytical algorithm, though precise (up to the numerical errors) only for the simplified viscosity, is actually very close and physically relevant to the Stokesian one.

Also presented in the lowest row of Fig. 7 are the density isosurfaces and slices, for which, just as in the figures for spectra, some regions are circled out to highlight the tiny differences in the patterns. Such differences from numerical schemes or physical viscosity models are negligible for both verification of our algorithm and discussions of our results, such as the general features of RSFs and the specific effects of helicity.

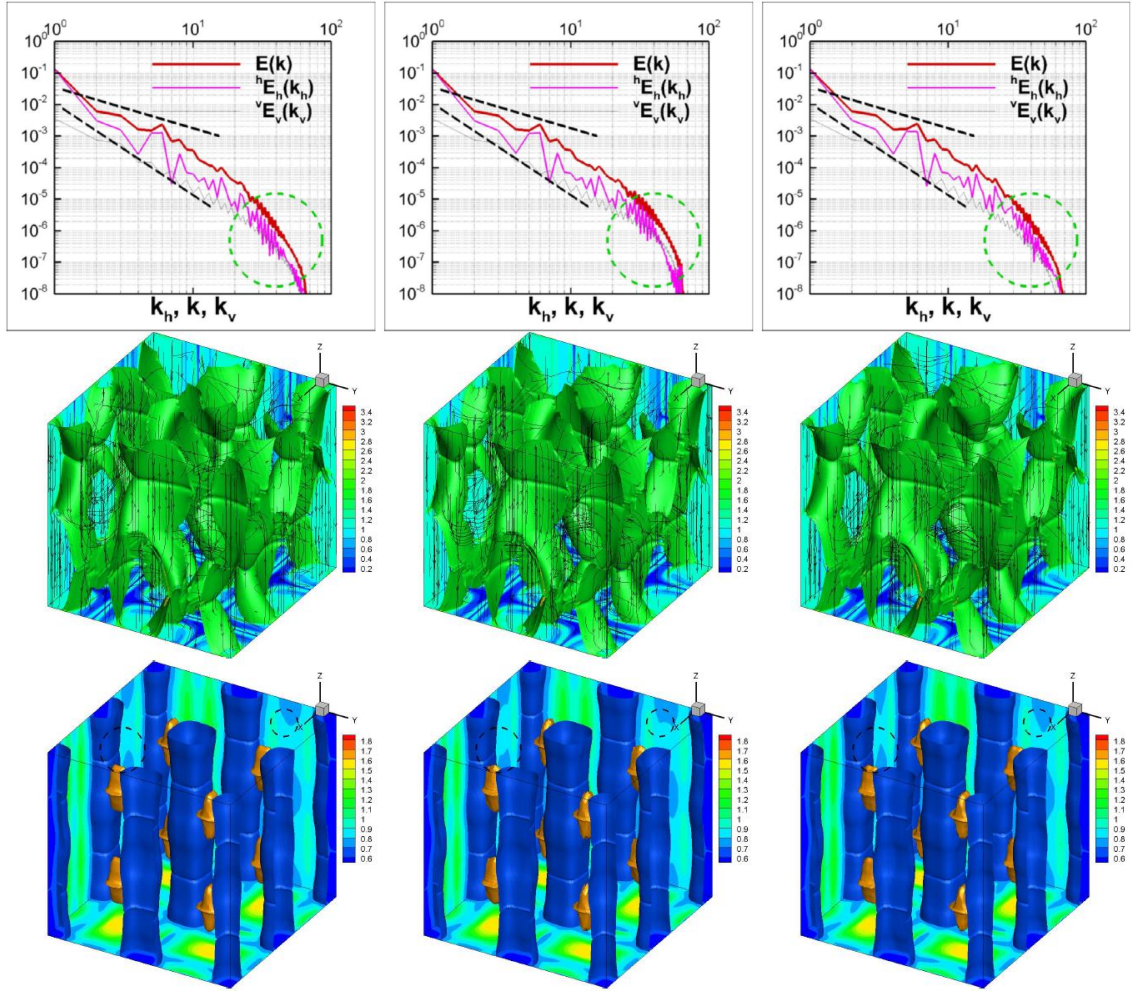

FIG. 7. Slices and isosurfaces (of values 0.7 and 1.7) of density (lowest row) at $t = 5$, showing tiny differences from different numerical schemes. Actually, the WENO-SYMBO (left column) produces maximum 0.532881 and minimum 1.82369, while WENO-Z (middle column) 0.532842 and 1.81667, a difference smaller than 0.4%. We also present the corresponding result from the Stokesian model (right column) simulated with the same strategy and algorithm described in Sec II C and the WENO-SYMBO finite difference scheme, and, again, only very small difference can be seen, with the maximum 0.53096 and minimum 1.82017, roughly of errors 2% ~ 4%. Dashed lines denoting for $k^{-5/3}$ (shallower) and $k^{-3}$ (steeper) laws in the (top-row) panels are also plotted for references, and some cirles are used to highlighted the regions where one can see the small differences. The middle-row panels are for the corresponding vorticity-amplitude slicies and isosurfaces with vorticity streamtraces, showing very close patterns.

Complex singularity structures however must be sensitive to the differences in the dissipation range, because the information of the singularities are presumably precisely reflected in the differences of such fashion. Thus, even though given a certain scheme or method, the differences in the results helical and nonhelical cases may indeed expose some fundamental aspects of the helicity effect, very careful analysis would be required with extra insights and techniques, which is beyond the scope here.

Before going into the even more specific discussions on the helicity effects on RSF multi-scale excitations of RSF eddies, we remark that although the RSF $E(k)$ appears to also maturize into a (approximately) $k^{-5/3}$ scaling law (designated with a slashed line) in the potentially inertial range, the vertical spectrum of vertical velocity $^vE_v(k_v)$ does not. The latter appearing to be $k^{-3}$ (also designated with a slashed line), as will be further addressed.

## B. Multi-scale excitations of RSF eddies from 3C2D initial fields

We now turn to compare results from the simulations started from the nonhelical and purely helical (actually Beltramian) RSFs described by Eqs. (18), both assisted by (20) with $\epsilon = 0.35$. The viscosity model satisfies Eq. (6).

Several necessary mathematical and theoretical elements for the presentation and explanation are in order. First of all, our compressible 2C2Dcw1C3D RSF velocity is subject to the general (refined) Helmholtz decomposition into right- and left-handed transversal components, and, the parallel one (Ref. 10 and references therein)

$$\boldsymbol{u} = {}^R\boldsymbol{u} + {}^L\boldsymbol{u} + {}^P\boldsymbol{u} \qquad (25)$$

in which the incompressible component $^I\boldsymbol{u} = {}^R\boldsymbol{u} + {}^L\boldsymbol{u}$ must be purely 2D in our cyclic box[26], the latter 2D requirement is essentially the Theorem 1. Also, Eq. (19) for an incompressible field should be written more precisely as

$$^R\hat{\boldsymbol{u}} = (\,^I\hat{\boldsymbol{u}} + \hat{i}\boldsymbol{k} \times {}^I\hat{\boldsymbol{u}}/k)/2 \text{ and } {}^L\hat{\boldsymbol{u}} = (\,^I\hat{\boldsymbol{u}} - \hat{i}\boldsymbol{k} \times {}^I\hat{\boldsymbol{u}}/k)/2 \qquad (26)$$

which correspond purely helical fields as in Definition 1 with respectively positive and negative helicity. The different components are mutually orthogonal, thus the inner product of $\boldsymbol{u}$ and the correpsonding power spectrum can be accordingly decomposed, respectively, into the sum of that of each component.

## 1. The general spectral dynamics

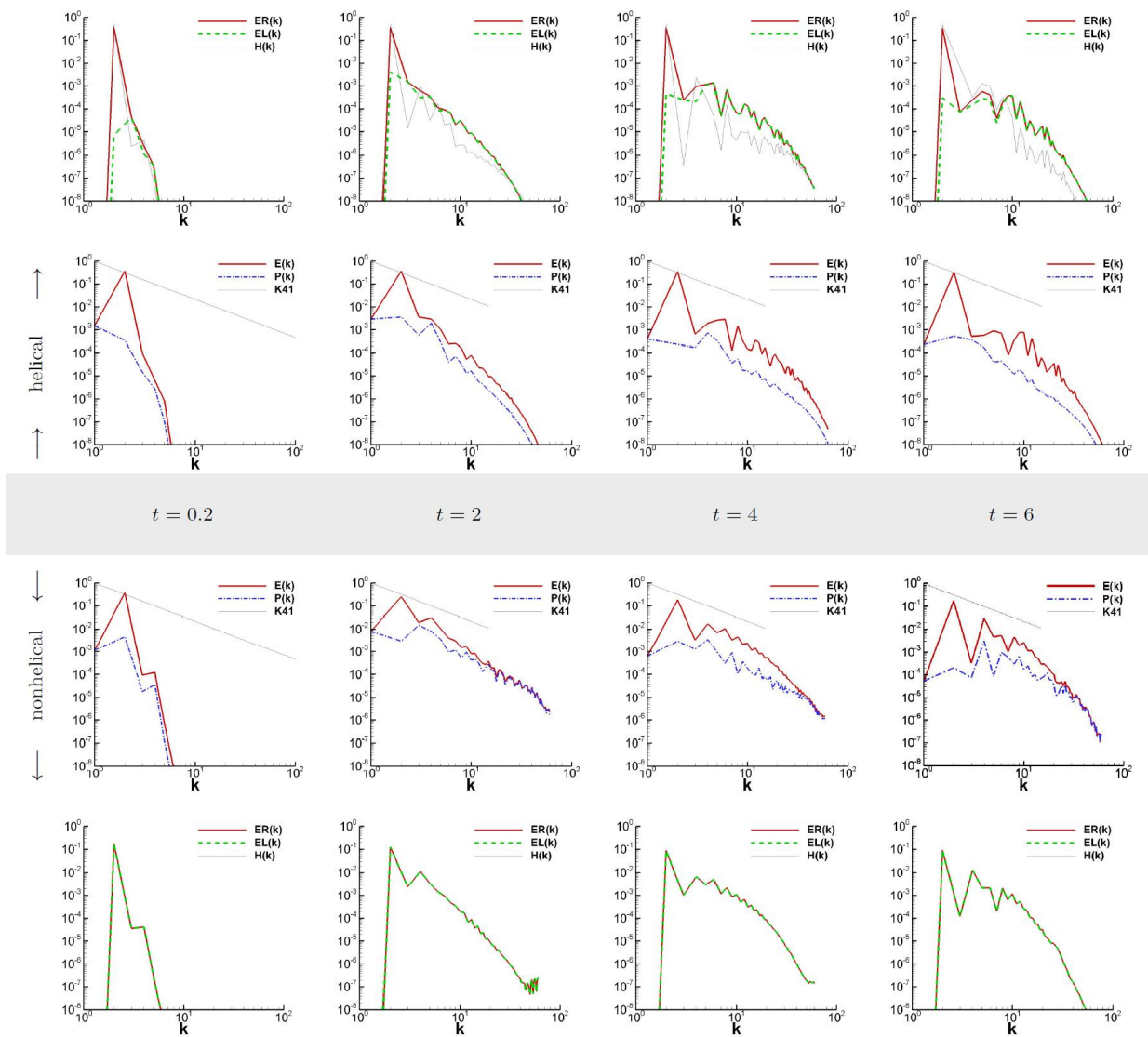

FIG. 8. The velocity, its left-, right-handed and parallel-mode power spectra $E(k)$, $EL(k)$, $ER(k)$ and $P(k)$ at different times for the helical and nonhelical cases, together with the Kolmogorov scaling law ("K41") for reference in some plots.

We should first point out that, since our initial RSF has wavenumber $k = \sqrt{2}$, counting it on the first or second integer shell, and similarly other modes for other shells, for plotting the spectrum presents different results. Thus, when interpreting our results, especially for small $k$, we should always be careful about this, which is the reason we have plotted the results with two different ways, one in Fig. 8 and others in Sec. V for the accelerated cases with such modes on the second shell (and similarly for others for which the issue is less critical with larger $k$) and the other with these modes counted on the first shell (for all the other plots in the work).

Since the field is composed of the left-, right- and nill-handed modes, i.e., $^L\hat{\bm{u}}$, $^R\hat{\bm{u}}$ and $^P\hat{\bm{u}}$ in the Fourier-space Helmholtz decomposition $\hat{\bm{u}} = {^I\hat{\bm{u}}} + {^P\hat{\bm{u}}}$ with $^I\hat{\bm{u}} = {^R\hat{\bm{u}}} + {^L\hat{\bm{u}}}$ (Ref. 10 and references therein), it is natural to look first at the spectral behaviors of such sectors in Fig. 8, for which we define according to Eq. (19), now for the transversal/incompressible component $^I\hat{\bm{u}}$, the power spectra of the left- and righ-handed sectors of the velocity, respectively,

$$EL := \sum_{|\bm{k}|=k} |^L\hat{\bm{u}}(\bm{k})|^2/2 \text{ and } ER := \sum_{|\bm{k}|=k} |^R\hat{\bm{u}}(\bm{k})|^2/2, \qquad (27)$$

and compute the spectra of $^I\hat{\bm{u}}$ and helicity, respectively,

$$P(k) := \sum_{|\bm{k}|=k} |^P\hat{\bm{u}}(\bm{k})|^2/2 \text{ and } H(k) := \sum_{|\bm{k}|=k} |\hat{\bm{\omega}}(\bm{k}) \cdot \hat{\bm{u}}(-\bm{k})|^2/2 = k(ER - EL). \qquad (28)$$

Obviously, $E = P + EL + ER$ from the orthogonality among the corresponding modes. The two rows above the shaded box containing time labels for each column are for the results from helical RSF, and those below for the nonhelical case. The total spectrum $E(k)$ and the K41 $k^{-5/3}$ law are also plotted for reference. Note that in the last row the helicity spectra are zero (up to the numerical errors) and not visible in the windows of the layouts.

We observe *i)* the early excitation of modes by the initial ones ($t = 0.2$), *ii)* most available modes being well excited but still far from the "equilibrium" ($t = 2$: the velocity spectrum is approaching but obviously not close to the K41 law), *iii)* premature "equilibrium" state with inertial and dissipation ranges almost but not yet established very well ($t = 4$) and *iv)* the mature equilibrium state with all excitation established as possible and the dissipation having been systematically reducing the whole level of spectrum ($t = 6$). After $t = 6$, the system is going towards the so-called 'late-time' decaying regime which is not of our interest here.

It is seen in Fig. 8 that the right-handed initial RSF evolves into that with all three sectors, but $ER$ keeps dominating the energetic behavior, especially before entering the dissipation scales (where the amplitudes of left- and right-handed modes are relatively closer), with $H(k)$ being positive definite [$H(k)$ could be very small and negative for some $k$s with $ER \approx EL$, which however does not change the overall feature]; while, the nill-handed ('racemic') initial RSF keep nill-handed (thus nonhelical), never loosing the symmetry as time goes.

Interestingly, the parallel modes of the helical case are less excited, both in the absolute and relative senses, the ratio $P(k)/E(k)$ for the relative value being simply measured by the vertical distance between the two spectral lines in the logarithmic coordinates for the plots [easily observable with bare eyes by comparing the areas between the lines of $P(k)$ and $E(k)$], except for the first two shells at $t = 6$, which is consistent with the mechanical/geometrical and statistical analyses, and, the 'fastening' notion proposed earlier[10]: the gravest modes of $k = \sqrt{2}$ counted on the second shell serve as the 'source', of energy and helicity, for the other modes, thus the time variations of the modes on it and the nearest shells can not be 'inertial'.

Since RSF is apparently anisotropic, the above general 1D spectral analysis is of course not complete, as already indicated in Sec. IV A, and more specific spectral results will be analyzed in Sec. IV B 3 before which we would like to present the general patterns of such RSF evolutions.

### 2. The general flow patterns

Fig. 9 presents the evolutions of density contours for the helical and nonhelical RSFs, showing consistently, with the previous spectral observation, especially the compressibility-relevant parallel-mode spectra, that, over all, the density-modes are less excited. Especially, the nonhelical case can have the maximum and minimum values far beyond the initial ones, indicating the production of strong expansion regions and compression (probably shocks), which will be further examined:

We now focus on the moment at $t = 6$, when the snapshots of the isosurfaces and slices of $\rho$ and $|\nabla \rho|$ are captured in Fig 10: the high(est) density regions are in general accompanied with the stong(est) gradients at almost, but not exactly, the same places, thus demonstrating the shocks with the isothermal density being essentially the pressure. The nonhelical case presents much stronger shocks with much more sharply 2D structures. In Fig. 11, we zoom in close to the strong gradient regions, rotate the axis for better observation, and plot the rakes of velocity streamtraces, showing that the nonhelical case indeed presents (almost) 2D surface of shock with abrupt deflection of the streamlines across it and that the helical case prsents a much thicker spatial domain accrossing which the streamlines, also oblique, do not present any obvious deflection (thus not close to a real shock). The shocks are clearly seen to be much less space occupying, an indication of (strong) spatial intermittency, and most of the space are occupied by expansion regions with density close to the background value and low weak gradients. The stronger/more intermittent the shocks are, the more space weak $|\nabla \rho|$ occupies. Note that, though rationally natural, the conjecture and remarks relevant to shocks in Ref. 10 went much beyond the arguments supporting the analysis made there: the analytical tractability of the statistical mechanical calculation required weak excitation assumption and that of geometrical analysis took the inviscid limit. Thus, the results about shocks presented here is not only an obvious support but also a strong indication of a more systematic theory behind it.

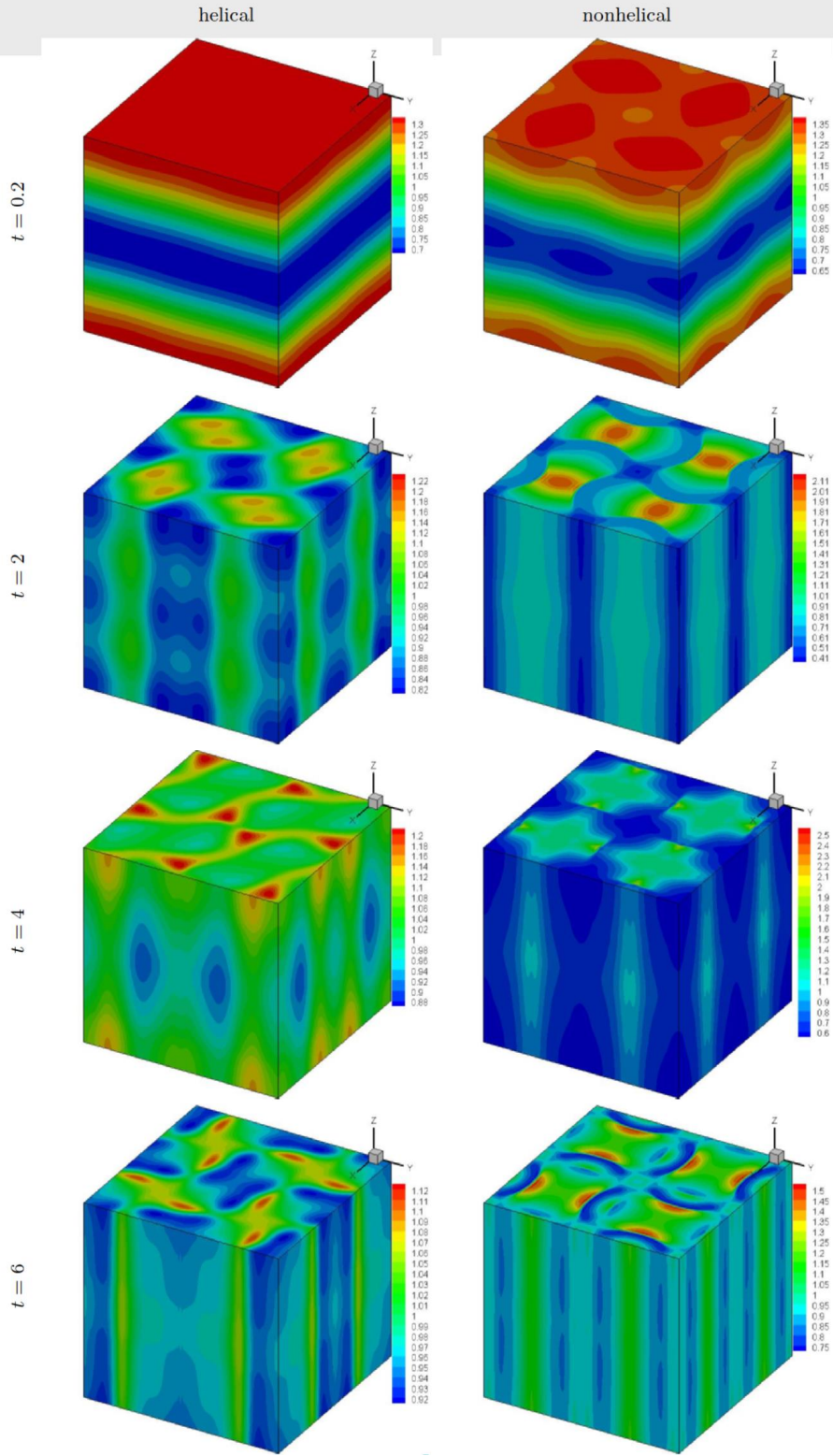

FIG. 9. The density contours of the helical and nonhelical cases at different times.

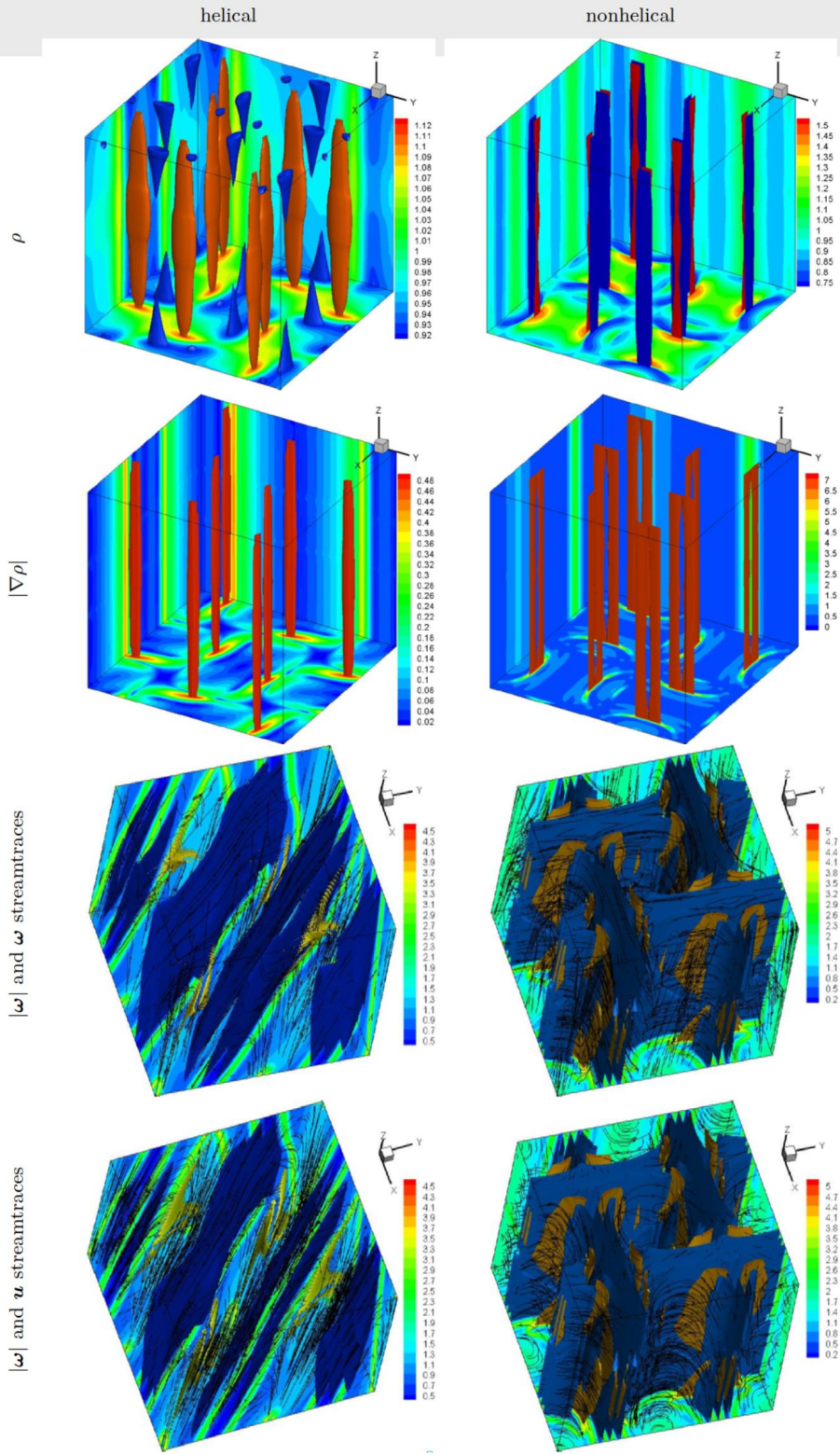

FIG. 10. Patterns of various quantities of helical and nonhelical cases at $t = 6$.

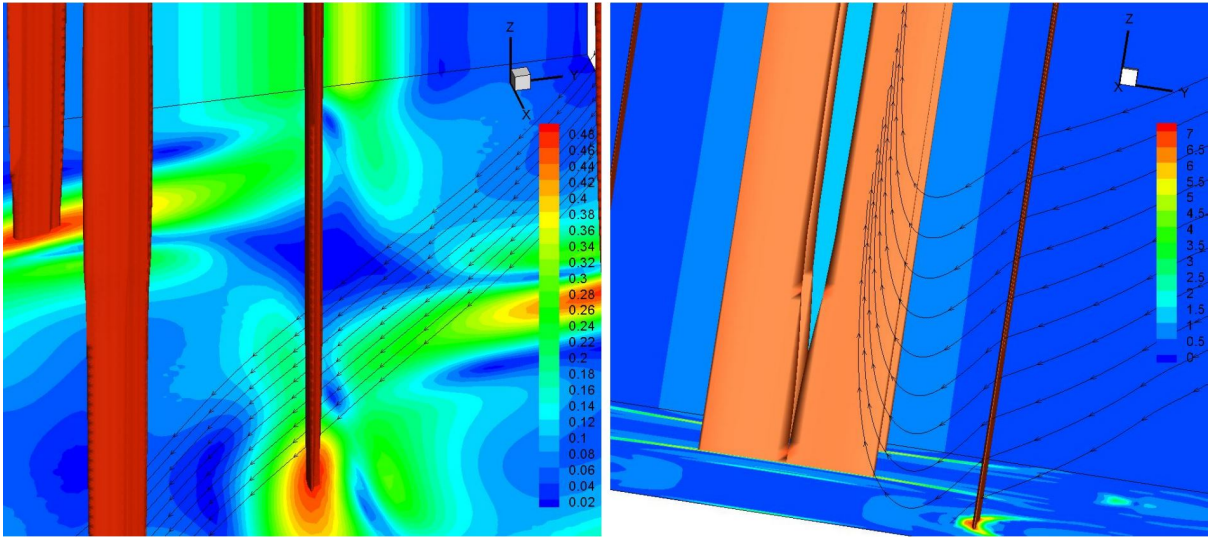

FIG. 11. Velocity Streamtraces across the strong density gradient regions for the helical (left) and nonhelical (right) cases.

We have explained in Sec. III A that the helical initial field is Beltramian, thus its $\boldsymbol{u}$- and $\boldsymbol{\omega}$-streamtrace patterns are exactly the same as presented there. Here, although, the helical case is not Beltramian, actually not even purely helical any more as time goes, the vorticity- and velocity-streamtrace patterns in the helical case, unlike the nonhelical one, look still quite similar in Fig. 10, at least at large scales, except for some small-scale details, which should not be surprising, because, as shown in Fig 8, at each moment, most of the helicity is concentrated on the $k = 2$ shell where the right-handed and left-hand energies are of a difference of (more than) three orders of magnitude, thus basically homochiral and Beltramian. The corresponding nonhelical patterns, in contrast, do not have such properties.

Note that, although RSF allows the $x_3$ dependence of $u_3$, which, however, in the classical setting is of the 'parallel mode' nature and in most spatial regions, as indicated by the above density discussion, should be small. Thus, RSF, if not particularly enhanced in the third dimension, appear over all much of 2D characteristic. One can start with an RSF of very large $u_{3,3}$, in which case $u_3$, compared to $\boldsymbol{u}_h$, will also be subjected to an additional damping operator $\nu\partial_{33}^2$ according to the viscosity model (6) and eventually will be weaker.

### *3. More specific analyses*

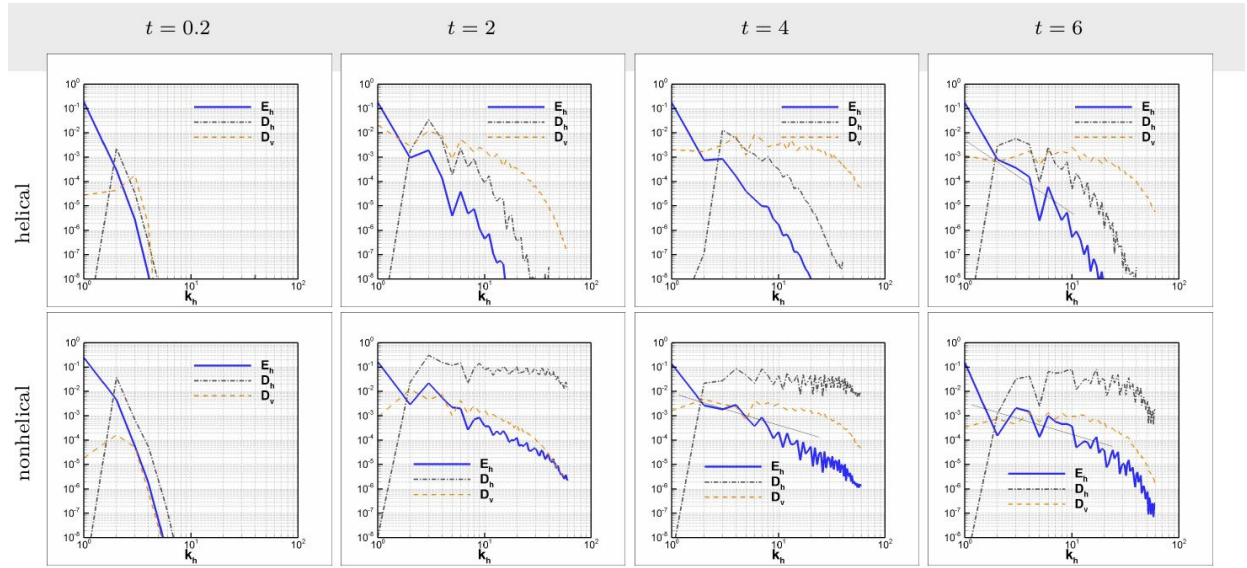

FIG. 12. The horizontal energy and divergence spectra, and, the vertical divergence spectra. $k^{-3}$ and the shallower $k^{-5/3}$ laws denoted by straight lines are also added to some plots for reference.

As remarked, RSF is by definition anisotropic, and the patterns of some quantities such as the $\boldsymbol{u}$ streamlines and those related to the density presented in Sec. IV B 2 appear to be quite of 2D character though in principle can be 3D. Thus, the general spectral dynamics presented in Sec. IV B 1 are not specific enough to describe RSFs, and, according to RSF nature and our initial data, we will look more into the details of the anisotropic properties.

First of all, since the RSF $\boldsymbol{u}_h$ is 2D, depending only on the horizontal wave vector $\boldsymbol{k}_h$, we should examine the relevant spectra in terms of $k_h$. Fig. 12 presents the 'horizontal spectra', corresponding to the results discussed in Secs. IV B 2 and IV B 1,

$$E_h(k_h) := \sum_{|\bm{k}_h|=k_h} |\hat{\bm{u}}_h(\bm{k}_h)|^2 \text{ and } D_h(k_h) := \sum_{|\bm{k}_h|=k_h} |\hat{\bm{u}}_h(\bm{k}_h) \cdot \bm{k}_h|^2, \qquad (29)$$

which measure the 'power' of the fluctuations of, respectively, $\bm{u}_h$ and $\nabla_h \cdot \bm{u}_h$ on each horizontal wave number $k_h$. [Note that the modes with $k_3 \neq 0$ for $\bm{u}_h$ are truncated by the definition of RSF.] Compared to the comparisons for the helical and nonhelical cases in Fig. 8, it is seen that here the $\bm{u}_h$ small eddies are even more markedly less excited: while $E(k)$ there nearly reaches a $\propto k^{-5/3}$ state at $t = 6$ in the 'potentially inertial' range (roughly for $2 < k < 10$), here $E_h(k_h)$ is much steeper, as can be seen by the thin solid black line denoting $\propto k^{-3}$ and added for reference. $D_h(k_h)$ also shows that horizontal small eddies are much less compressible. These horizontal behaviors are in sharp contrast with the nonhelical case where thin black lines denoting $\propto k^{-5/3}$ are added to the plots at $t = 4$ and $t = 6$. The vertical divergence spectrum

$$D_v(k_h) := \sum_{|\bm{k}_h|=k_h, k_3} k_3^2 |\hat{u}_3(\bm{k}_h, k_3)|^2 \qquad (30)$$

also appears to be slightly affected by the helicity, but much less than $E_h(k_h)$ and $D_h(k_h)$. Although such reduction of horizontal compressibility appears to be consistent with an argument given in Ref. 10, of boosting to a rotating frame in which the helicity vanishes and thus transforming the helicity effect to the rotation effect (Taylor-Proudman effect of reducing compressibility in the rotating plane), the additional detailed observations, such as the $E_h(k_h)$ spectral behavior, deserve further theoretical consideration according to the dynamics. [A consistent consideration of 'compressibility' involves normalization, say, by the enstrophy or total spectrum: c.f., the remark for Fig. 8 on the relative reduction of $P(k)$.] Here, we offer a possible theoretical scenario in the context of turbulence as follows.

The $\bm{u}_h$ dynamics in the isothermal RSF model used in our simulations is controlled by Eq. (1b-) which is almost autonomous except that the pressure gradient is coupled to the rest of the system through the density. Now, if the pressure gradient is assumed to be decomposed into two parts, one corresponding to that in the incompressible 2D flow (thus solving the Poisson equation with the source being the divergence of the nonlinear term) and the other being externally affected and serving as the pump for $\bm{u}_h$. If the pumping is concentrating at some large scales and the $\bm{u}_h$ is approximately incompressible (otherwise consider the transverse/incompressible component), a forward enstrophy transfer together with an $E_h$ spectrum $\propto k_h^{-3}$ (with logarithmic correction) is then the well-tested incompressible 2D turbulence theoretical result: depending on the compressibility which determines the partition of velocity excitations between transverse and longitudinal components, $E_h$ may present different behaviors, with the possibility of even a $k^{-5/3}$ law when the forwardly transfered 'energy' ($L^2$ norm) of the longitudinal component dominates in the strongly compressible situation, as seems to be the case of nonhelical results at $t = 6$ in Fig. 12. In other words, the rest of the system may offer an environment of external transfer channels while still facilitating a genuine internal transfers of the 2D incompressible flow nature. How Eq. (1b-) couples with the rest of the system can be very subtle due to the multi-scale and nonlocal nature of pressure/density fluctuations, and the characteristics at $t = 6$ in Fig. 12 for the helical case is of course not yet clear and clean enough; and, much more systematic investigations with general initial fields, including an ensemble of random ones, other than the specific one we treated here are needed to test such a speculation (but see also the results of driven/accelerated cases in Sec. V).

We then present in Fig. 13 the spectra of $u_v := u_3$, as functions of $k$, $k_h$ and $k_v$,

$$E_v(k) := \sum_{k=|\boldsymbol{k}|} \frac{|\hat{u}_v|^2}{2}, \quad {}^h E_v(k_h) := \sum_{k_3, k_h = |\boldsymbol{k}_h|} \frac{|\hat{u}_v|^2}{2} \text{ and } {}^v E_v(k_v) := \sum_{\boldsymbol{k}_h, |k_3| = k_v} \frac{|\hat{u}_v|^2}{2}, \quad (31)$$

which characterize the distributions and variations of, respectively, the vertical velocity, $u_v$, eddies in full 3D space, in the horizontal plane and along the vertical coordinate. We now see that the $u_3$ dynamics are not so much different for the helical and nonhelical cases as for $\boldsymbol{u}_h$. Especially, the ${}^v E_v(k_v)$ behaviors are very close, both presenting a quite clean $\propto k^{-3}$ law at large scales (thin lines of such a scaling are added to the plot for reference), and only at $t = 6$ the helical ${}^v E_v(k_v)$ is at a slightly higher level (while those of $E_v(k)$ and ${}^h E_v(k_h)$, both being close, are lower, beyond the first shell). Reference lines of $\propto k^{-5/3}$ are also added to the plots, denoting the possible (approximate) behaviors of $E_v(k)$ and ${}^h E_v(k_h)$ in the potential inertial range: just as before, doing this, we by no means indicate the precise results in the asymptotic inertial range, with or without corrections from intermittency or other physical reasons. [However, we should remark that the forward transfer of the enstrophy of $\boldsymbol{u}_h$ and the forward transfer of the power of $u_3$ (and even the parallel mode of $\boldsymbol{u}_h$) is not conflicting and can coexist along the same scales: see Sec. V for more relevant remarks on the driven/accelerated cases.] No strong theoretical suggestion, relevant to passive scalar or acoustics, is offered for the $k^{-3}$ scaling either. We believe these results, though coming only from the very specific initial fields, are fundamental and important, with no available directly relevant theory for explanation, to our best knowledge, and should be simply presented for information and motivation.

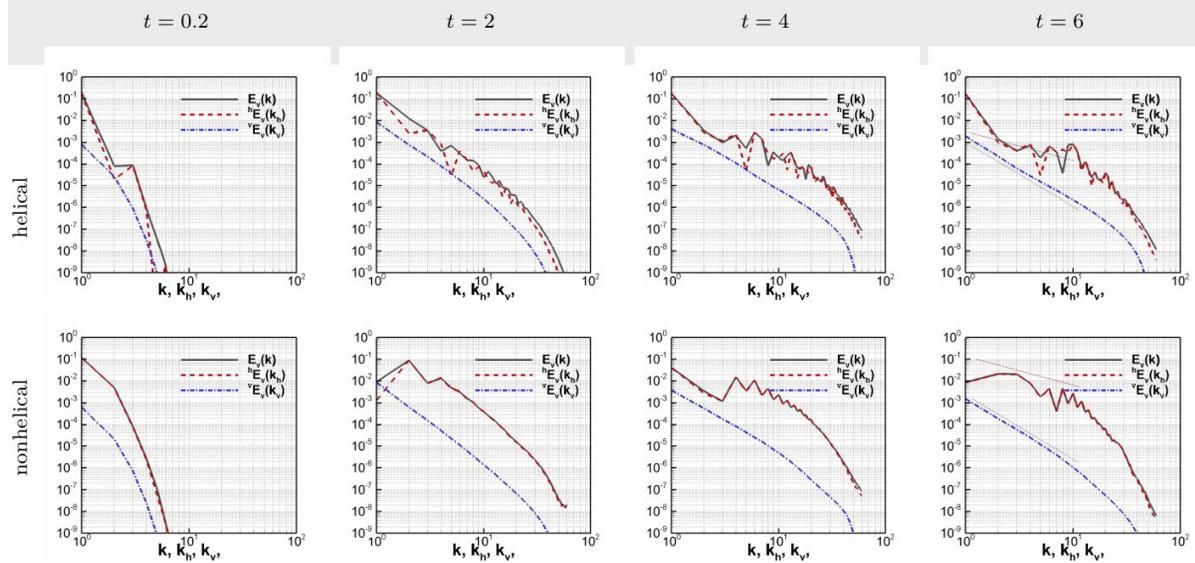

FIG. 13. The velocity spectra $E_v$, $^hE_v(k_h)$ and $^vE_v(k_v)$. Thin straight lines for laws $\propto k^{-5/3}$ (shallower) and $\propto k^{-3}$ (steeper) are added to some of the plots for reference.

## V. FURTHER DISCUSSIONS

We have already remarked that our algorithm also applies for the driven/accelerated cases, as long as the acceleration is of RSF, and we have also remarked that the 'fastening' notion of helical RSF (CBF) in Ref. 10 is supported. Thus, it would not be complete without showcasing the computation of driven/accelerated RSFs and exhibiting the relevant results, including those in the low-Mach-number regime with necessary physical discussions.

### A. Driven/accelerated RSFs

The accelerations for the nonhelical RSFs are precisely of the form of Eq. (17). We have already shown that the field of RSF can not be purely helical without being of 3C2D character, so combinations of Eqs. (17) and (18) were made to keep both chirality and three-dimensionality of the helical RSF accelerations. Computations of various combinations (including the variations with $a_3$ of different, even dominating, orders of magnitude)[24] and different Mach numbers were performed, but here we only focus on the closely relevant results for the purpose of this note. Now, as in Fig. 8 to emphasize the relevant physics, modes of $k = \sqrt{2}$ are counted on the second shell, i.e., $k = 2$, in all the plots.

### 1. $Ma = 1$

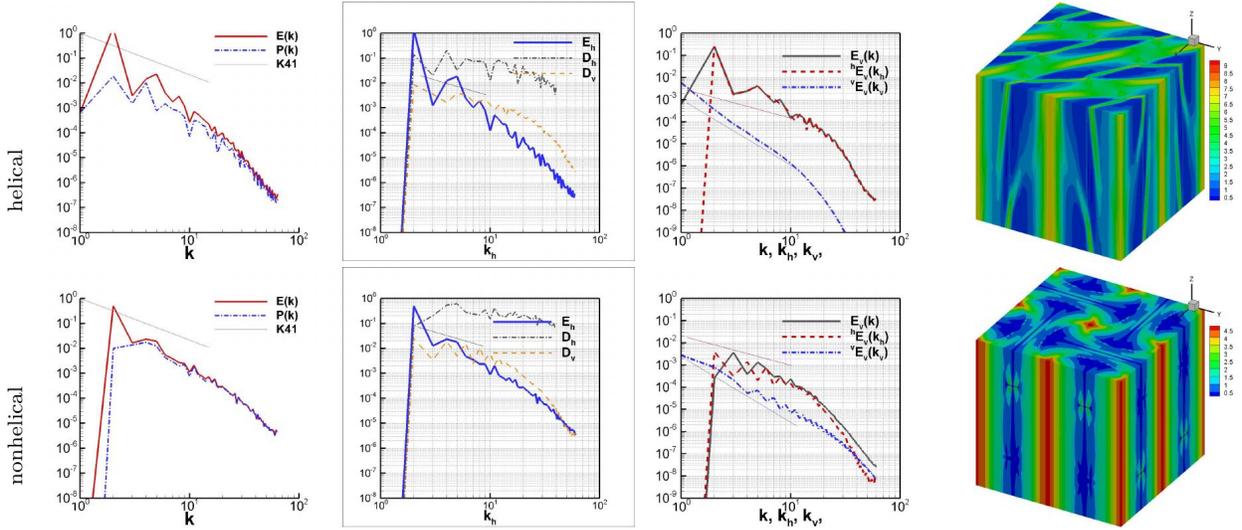

FIG. 14. The spectra and vorticity-amplitude-contour patterns for the helical (upper row) and nonhelical (lower row) cases of $Ma = 1$ at time $t = 10$, each column of panels corresponding, respectively, to those of Figs. 8, 12, 13 and 10.

The first 'nonhelical' and 'helical' computations for comparison were performed with accelerations, respectively, purely of the form of Eq. (17) and an equally weighted combination of Eqs. (17) and (18). Both initial fields were also given by Eq. (17), which however is irrelevant to the long time limit (but of course not supposed to be the case for all other models not studied here): such choice of the initial 'guess' of the solution was only to shorten the computation time. Long-time integrations reach the quasi-steady states (to be used for our discussions) with very stable spectra and flow patterns: absolutely steady states do not appear to be available within the costs we could afford but are neither necessary for our purpose, and, actually, so far we can not preclude the possibility of the existence of (quasi-)periodicity or some kind of recurrence in (some weak part of) the excitations.

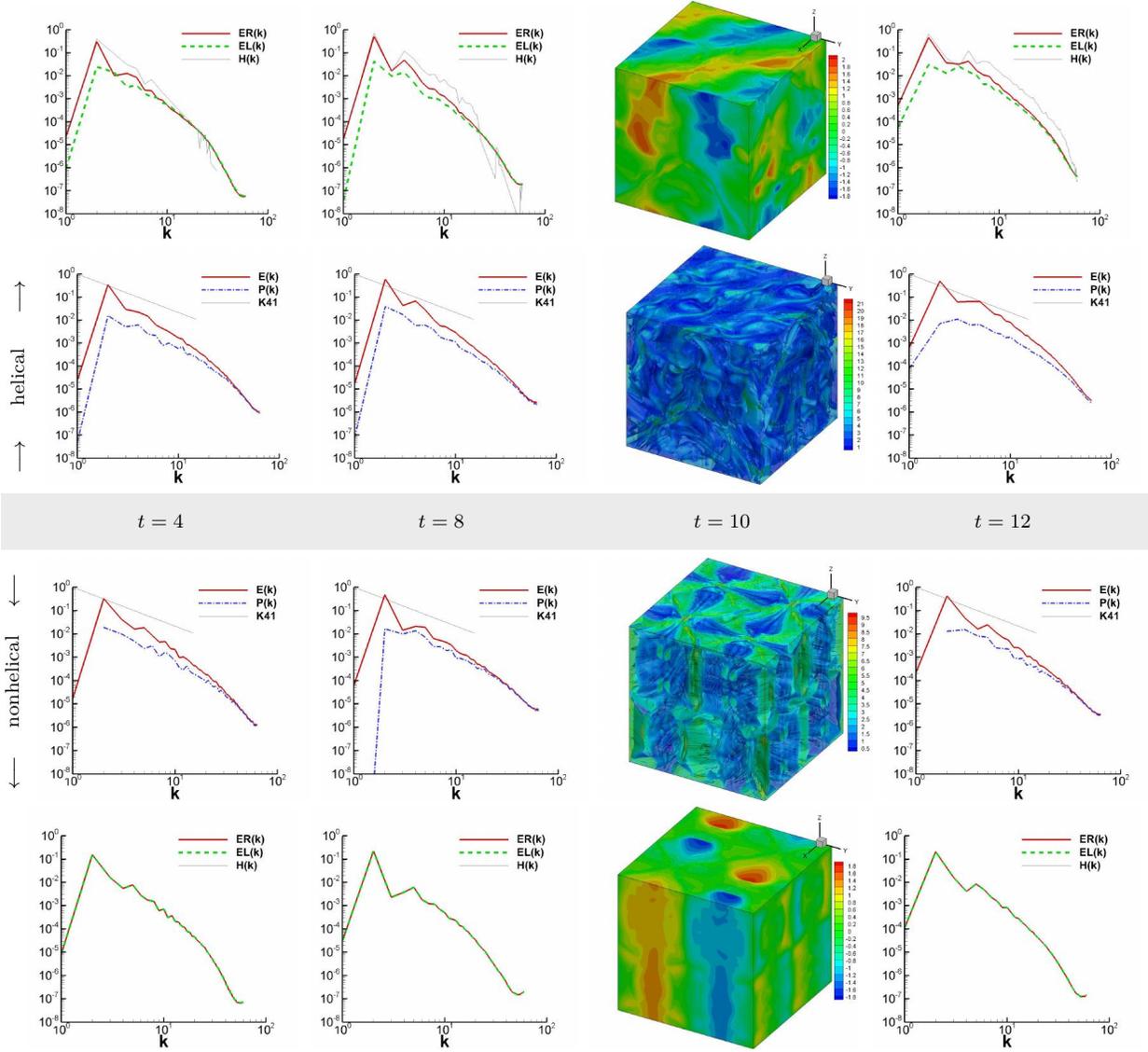

FIG. 15. The same NSF ($Ma = 1$) plots as Fig. 8, except that the plots with helical, i.e., right- and left-handed (top and bottom rows) and parallel (middle rows sandwiching the time-label box) spectra at $t = 10$ are so close to those at $t = 12$ that they are respectively replaced with the contours of $u_1$ (of fully 3D character now, unlike the 2D $u_1$ of RSF) and $|\boldsymbol{\omega}|$, the latter being translucent with velocity streamtraces.

Fig. 14 presents the corresponding (quasi-)steady state results (*c.f.* the caption for more information) to those relaxation/decaying ones in the previous sections, the same analysis further supporting the notion of 'fastening the flow with helicity'. The lines for scaling laws of $k^{-5/3}$ (shallower) and $k^{-3}$ (steeper) are just for references, with the previously discussed results and the conventional turbulence scenario in mind. The physical scenario and flow patterns will be further addressed below together with those of $Ma = 0.1$, before which we would like to remark that, as in the decaying case, the corresponding NSFs as presented in Fig. 15 with respectively the same accelerations (c.f. the caption for more information) evolve into rather different states, but, with similar analysis, consistently supporting the 'fastening' effect of helicity. An interesting point to remark on is that, in the nonhelical case, the parallel modes, characterized by $P(k)$, at $k = 1$ are barely excited as in RSF, even more so for NSF to the point that at some times (e.g., $t = 4$ and $t = 12$ in Fig. 15) $P(1) = 0$ in our double-precision computation and postprocessing of the data: more results and analysis about this issue are available[24], but a precise understanding from the mode interaction point of view is still missing, calling for systematic mathematical investigations.

### 2.  $Ma = 0.1$

We now turn to the cases in the low Mach number regime. The setup for comparison is the same as that in Sec. V A 1, except that $Ma = 0.1$. Fig. 16 presents the corresponding spectra that we have discussed in the previous sections, showing that the 'fastening' notion is also valid.

The flow patterns are reminiscent of the so-called 'vortex crystal'[50,51] of 2D flows (c.f., the topviews, left and right panels of Fig. 17, of the same plots of respectively helical and nonhelical patterns of density with velocity streamstraces in Fig. 16), and of the template of Jiu-Gong/Ba-Gua (or nigh-palace/eight-trigram) in the ancient Chinese astronomy and philosophy (about, say, "change" and "balance"[52]): here, each 'palace' swirls, helical or not, and, if the domain being considered realistically as a torus with no boundary and only the horizontal plane being considered, the two pairs of opposite 'borders' should be 'patched' and the algebraic sum of the number of full cyclones are four. However, we can also think about tiling the space with boxes of periodic boundaries. In the latter case, the algebraic average number of cyclones in each box is still four,[53] but, now, again with only the horizontal plane being considered, geometrically, the quarters in the four corners or halves at the four centers of boundaries belong to different cyclones pointing to eight directions from the central one, which is precisely the meaning of the Jiu-Gong/Ba-Gua pattern. That is, there are always eight cyclones encircling the central one (totally nigh). It is interesting that now such patterns are also observed in the cyclone cluster surrounding Jupiter's northern polar ('octagonal' or 'ditetragonal'[27,28]). The 'quasi-pentagonal shape' of cyclones surrounding Jupiter's southern polar have been claimed to be reproduced and analyzed in more realistic models (shallow-water[54] and full Navier-Stokes with deep convection and Coriolis force[55]) than the pure 2D incomplressible Euler or Navier-Stokes[51,60], but not for such northern octagonal pattern, and it is possible to find between our results and Jupiter's cyclones deeper mathematical and physical connections the details of which however will be discussed elsewhere[24]: the cyclic boxes being not supposed to be the realistic building blocks in a "Lego" set for Jupiter's atmosphere, fitting the physical situations (e.g., Refs. 54 and 55) to model Jupiter's cyclones is not the purpose here, and pointing this out is just to emphasize the natural bridging of 2D and 3D dynamics by RSF.

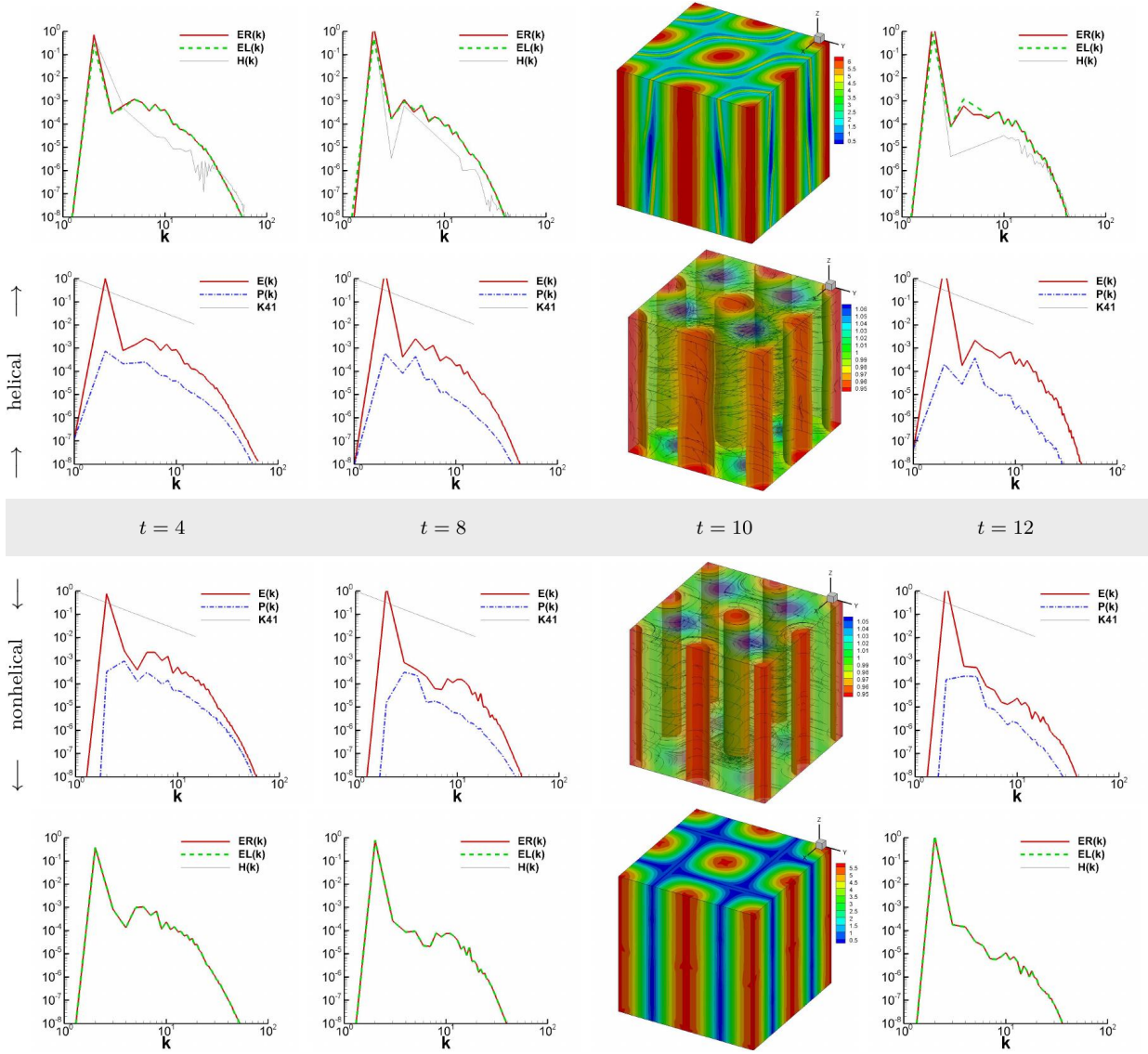

FIG. 16. The same RSF ($Ma = 0.1$) plots as Fig. 15, except that the helical and parallel spectra at $t = 10$ in the quasi-steady state are so close to those at $t = 12$ that they are respectively replaced with the contours of $|\boldsymbol{\omega}|$ in the top and bottom rows and $\rho$ (translucent and with velocity streamtraces) in the middle rows sandwiching the time-label box.

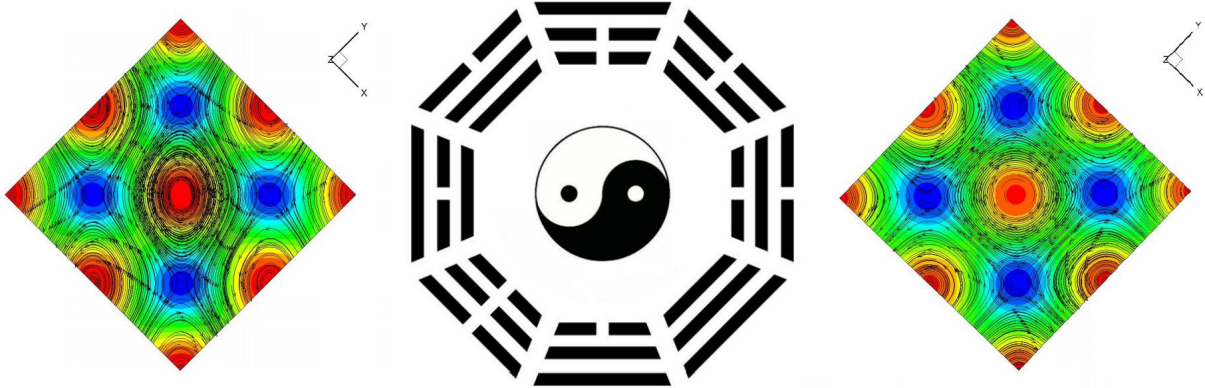

FIG. 17. Topviews (left and right panels) of the same plots of the density contours with velocity streamtraces as in Fig. 16. The middle panel is a Jiu-Gong/Ba-Gua (or nigh-palace/eight-trigram) template whose pattern the RSF ones on the two sides resemble.

Comparing the patterns of density and vorticity between the helical and nonhelical cases in our calculations, we see that the density patterns appear to be more 'stable' than that of vorticity-amplitude ones, in the sense that the 'nigh palaces' of former are more of circular shapes with clear boundaries inbetween, not as much affected by helicity as the latter: In the helical case, the vortices are not well shielded by vanishing vorticity zones, and they are stretched to be reminiscent of the 'ovals' (say, the red, white and others of Jupiter[56]): while the ovals in Jupiter may be stretched by zonal flows, the relevant effective ingredient of the latter might be related to the helicity. At this point, we do not have a complete understanding of this issue but can only conjecture that it may be due to the RSF regularization in Eq. (1a) by the spatial averages in the right hand side (r.h.s.): note that each term in the r.h.s. is the r.h.s. of time variation rate of the corresponding spatial average of $\ln \rho$, a very 'strong' regularization.

Relevant incompressible two dimensional turbulence theories and arguments date back to the work of Onsager[57], Lee[58], Fjørtoft[59] and Kraichnan[19], and, also recent ones respecting all the ideal invariants and involving regional maximum entropy calculations[60], with particularly the notice of the subtleties in setting up the forcing scheme for the 'vortex crystal'[51]. Our finding, on the one hand, appears on the first sight somewhat more trivial in the sense that the nigh-palace/eight-trigram pattern is mainly the accumulation of 'energy' at the scale where it is accelerated, but on the other hand, is obviously much more nontrivial that, among others, the RSF model is compressible, coupling with the 3D density and vertical velocity dynamics, but still not the full Navier-Stokes, particularly no explicit Coriolis acceleration (but see the connection of RSF with rotation remarked in the introductory discussions): actually, even the first point on the accumulation and then concentration of the 'energy' at the accelerated wavenumber is highly nontrivial, in the sense that now $k_h = 1$ ($\boldsymbol{k}_h = \{\pm 1, 0\}$ and $\boldsymbol{k}_h = \{0, \pm 1\}$) modes of $\boldsymbol{u}_h$ are basically not excited (while for other schemes not shown here such $k_h = 1$ modes can be excited to an even higher level of energy than the accelerated modes, but then with no Jiu-Gong/Ba-Gua pattern resembling that of Jupiter, simply by the intrinsic inverse transfer mechanism in the scenario to be discussed below), which, from many numerical experiments, some of which also presents similar but less clean Jiu-Gong/Ba-Gua pattern with accelerations of the same fashion but added to $k_h > 1$, is found to be associated to a similarity law of (quasi-)steady state concentration and accelerations, i.e., the 'acceleration-concentraton similarity law' (ACSL),

**Claim 1** *(ACSL) When the acceleration is added at $\boldsymbol{k}_h^a = \{k_1^a, k_2^a\} \neq \boldsymbol{0}$, the power spectrum $E_h(k_h)$ peaks either at $k_h = 1$ or $k_h = \sqrt{2}$ and the ratio $E_h(1)/E_h(\sqrt{2})$ is a monotonically increasing function of $s_a := (|k_1^a| - |k_2^a|)^2/(|k_1^a|^2 + |k_2^a|^2)$, valid for both 2C2Dcw1C3D RSF and 2D NSF in boxes of $2\pi$ periodicities.*

The singular $\boldsymbol{k}_h^a$ in such ACSL, so far empirical and probably not having been precisely described (for missing some details such as the preconditions behind the finite numerical experiments), is far from complete in describing wavevector distributions of accelerations. For example, modes of more than one $\boldsymbol{k}_h^a$ can be simultaneously accelerated, which may also be associated with some similar but much more complicated law whose description needs parameters beyond the above $s_a$ and will not be discussed here. As just mentioned before the above Claim 1, also an emperical *law* in our experiments *in silico* is that the smaller scales the same-fashion accelerations as we have applied here, with $s_a = 0$, are acting on, the lower degrees of condensations of $E_h$ at $k_h = \sqrt{2}$ and thus less clean are the octogonal/ditetragonal patterns; similarly is the condensation of $E_h$ at $k_h = 1$, presenting no Jiu-Gong/Ba-Gua pattern of cyclones, with $s_a = 1$. Then, the trivial corollary is that when the accelerations are exerted right on modes of $k_h = 1$ or $k_h = \sqrt{2}$, most of the energy is just accumulated there (by the intrinsic condensation mechanism — see below), with coexisting forward transfers of the enstrophy (of $\boldsymbol{u}_h$) and compressive-mode energy. Thus, the numerical results we have presented here correspond to the most interesting (at least in the sense of cyclone pattern resembling that of Jupiter) part of the *corollary*. Our 2D NSF experiments (not shown) with small Mach number (Ma = 0.1) indicates that ACSL presumably also works in incompressible 2D case. Intuitively, the ACSL should be related to some kind of 'resonance' between the forcing and the small-$k$ dynamics, which might be the clue for a systematic mathematical analysis. We have to admit that, without presenting all numerical results of various cases, the above descriptions of the ACSL are still not clear enough, but we hope it does be sufficient to deliver the general idea of a much bigger picture by putting our current results into the background of more general acceleration schemes acting at various scales in different ways. Tremendous details will be communicated elsewhere to specifically address this issue[24], but we would like to further remark that, if our RSF model dynamics are indeed physically underlying the corresponding cyclone cluster on Jupiter, our results then further indicate additional information, associated to the ACSL, on ingredients such as the storms, driving the cyclones. [As will be indicated in the discussions of the next paragraph, the horizontal compressibility of our 2C2Dcw1C3D RSF does not appear to be essential for such results, thus, according to the relevance with the formal fast rotating limit of compressible NSF of the horizontally incompressible 2C2Dcw1C3D RSF, the physical relevance with Jupiter's atmosphere does seem to be favorable. However, the fundamental connection, if indeed, would become evident only if the empirical ACSL is systematically understood or proven to be a 'theorem'.] And, such concentration of energy at $k_h = \sqrt{2}$ to a level much higher (one to several orders of magnitude larger) than any other shells is supported by the following RSF spectral transfer scenario.

The scenario of forward enstrophy transfer of $\boldsymbol{u}_h$ coexisting with the forward transfer of the power of $u_3$ and the parallel mode of $\boldsymbol{u}_h$ (as indicated by the lines designating the $k^{-5/3}$ law in various plots) in the (potentially) inertial regime, as suggested in Sec. IV B 3, is consistent with all the results presented here (note that the dominance of $E_v$ in the regime designated by the $k^{-5/3}$ reference lines is implied in the varioius plots of spectra, see also those to be presented below). Note that the energy-flux-loop scenario together with the numerical analysis of dual cascades for 2D compressible turbulence, resembling that in a stably stratified 2D incompressible turbulence[17], has been offered by Falkovich and Kritsuk[15], and the possibility of additional 3D ('fast/wave') channels beyond the 2D ('vortex/slow') horizontal dynamics in incompressible rotating flows facilitating some specific 2D excitations was also proposed in Ref. 26 with relevant transfer results reported in Ref. 16. Now, it appears that a combination of such ideas, though neither so far has been founded by firm dynamical theory, is working for our compressible RSFs. Note also that we have only discussed the 'power' (spectrum) of velocity (component), not particularly the 'kinetic energy' weighted by the density, the latter, physically important and useful in many aspects though (e.g., Ref. 15), has not as clear and direct connections with the flow patterns presented here, to our point of view. It appears favorable to assume the 'cascades' of such $L^2$-norm quantities in the inertial range of fully developed compressible RSF and NSF turbulence, as our results preliminarily indicate.

Helmholtz decomposition of the velocity into transverse and longitudinal components can be applied to obtain the respective dynamical equations in which the vortex-acoustic interactions terms present and may be treated perturbatively to some degree for weak excitations[21,61]. It might be possible to carry the ideas further to analyze the spectral transfer issue, however, our conclusions also work in strongly compressible RSFs with shocks in Sec. V A 1, calling for more complete and systematic analysis. Here, we offer further results for information:

*a. The pressure term:* The horizontal pumping from the pressure term, the only 'active feedback' from the system beyond the $\boldsymbol{u}_h$ equation, in the scenario suggested in Sec. IV B 3 and further discussed in the above, has been assumed to be concentrating at the large scales, which is now checked to be indeed the case: according to Eqs. (3 and 4) for the pressure gradient, we present in Fig. 18 the following spectra associated to the gradient of $\zeta := \ln \rho$

$$E_\zeta(k) := \sum_{k=|\boldsymbol{k}|} \frac{|k\hat{\zeta}|^2}{2}, \ ^h E_\zeta(k_h) := \sum_{k_3,k_h=|\boldsymbol{k}_h|} \frac{|k_h\hat{\zeta}|^2}{2} \text{ and } \ ^v E_\zeta(k_v) := \sum_{\boldsymbol{k}_h,|k_3|=k_v} \frac{|k_v\hat{\zeta}|^2}{2}, \quad (32)$$

showing also clearly that most of the gradient is contributed by the horizontal variations.

*b. Each velocity component:* Fig. 19 leaves little 'magic' in the $k^{-5/3}$ scaling law in the $E(k)$ presented in various figures earlier: we have also computed the parallel/longitudinal-

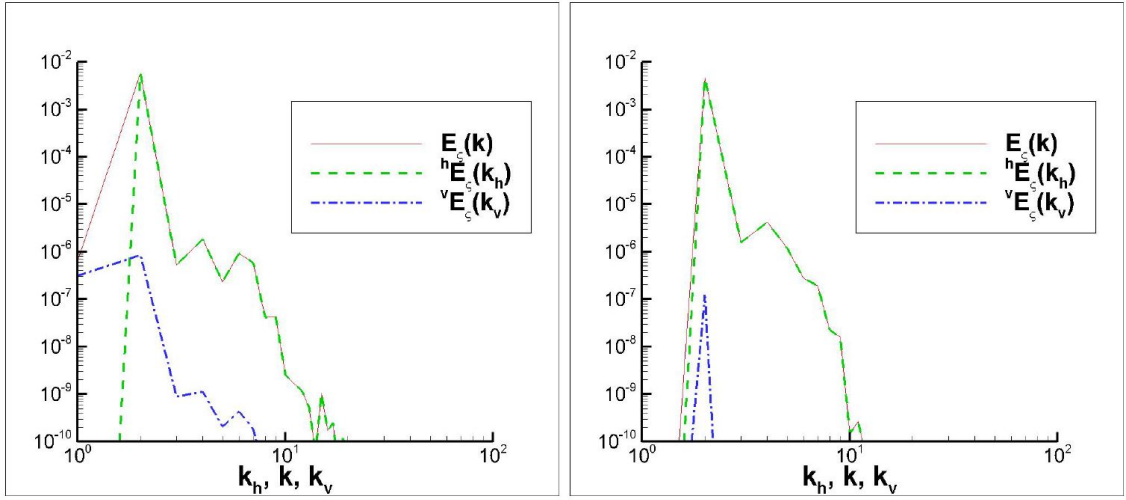

FIG. 18. The pressure-term spectra at $t = 12$ for helical (left) and nonhelical (right) cases; Ma=0.1

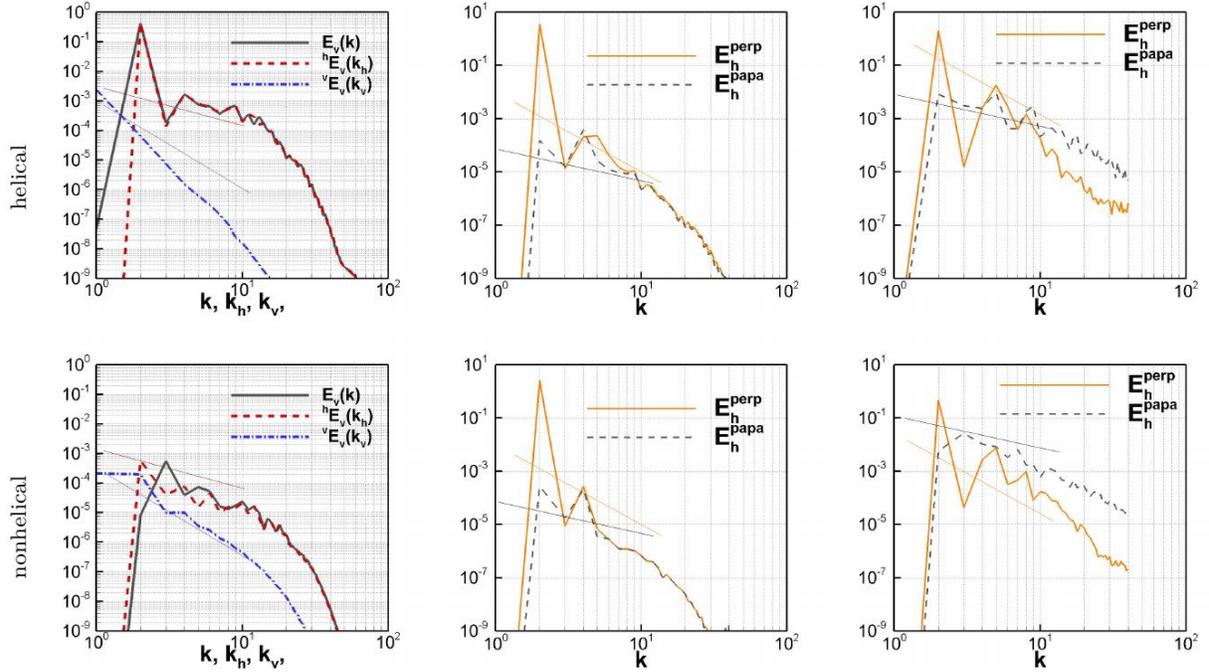

FIG. 19. The vertical velocity spectra ($Ma = 0.1$: left), $E_v$, $^hE_v(k_h)$ and $^vE_v(k_v)$ as in Fig. 13, and the parallel/longitudinal- and transversal-mode spectra of $\boldsymbol{u}_h$ for $Ma = 0.1$ (middle) and $Ma = 1$ (right). Thin straight lines for laws $\propto k^{-5/3}$ (shallower) and $\propto k^{-3}$ (steeper) are added to some of the plots for reference.

and transversal-mode spectra of $\boldsymbol{u}_h$

$$E_h^{para}(k_h) := \sum_{|\boldsymbol{k}_h|=k_h} |\hat{\boldsymbol{u}}_h \cdot \boldsymbol{k}_h|^2/k_h^2; \; E_h^{perp}(k_h) := \sum_{|\boldsymbol{k}_h|=k_h} |\hat{\boldsymbol{u}}_h - \boldsymbol{k}_h \hat{\boldsymbol{u}}_h \cdot \boldsymbol{k}_h/k_h|^2. \qquad (33)$$

Note that $\hat{\boldsymbol{u}}_h = 0$ for $k_3 \neq 0$, thus $k_h$ is essentially $k$ for whatever $E_h^\bullet$. We see that, for the case of $Ma = 0.1$, $E_v$ (actually $^hE_v$) dominates the $k^{-5/3}$ power spectra, thus presumably the forward spectral transfer, beyond $k=2$; $E_h^\bullet$ are of similar amplitude with $^\bullet E_v$ for $k<5$ but fall off quickly starting from $k=5$, which makes the spectral behaviors look somewhat unclear, but still we see that they are not inconsistent with the asymptotic $E_h^{para} \propto k^{-5/3}$ and $E_h^{perp} \propto k^{-3}$ corresponding respectively to the forward 'energy" and 'enstrophy' transfers from simple dimensional analysis[19,47]. The right panels of $Ma=1$ for comparison present much higher levels of $E_h^\bullet$ and the asymptotic scaling laws appear to be clearer: also clearer is the (relative) reduction of $E_h^{para}$ with helicity, i.e., the 'fastening' notion.

## B. Summary and outlook

RSF being new and anisotropic, many more results can be still of some value to be demonstrated. For example, we may analyze the vertical and horizontal helicity, dissipation and density spectra, and, also the anisotropic higher-order structure functions of various variables. We may also analyze the longer (late) time evolution, such as the anisotropic decaying rates, the homogeneization processes, and so on and so forth. However, to make all that more meaningful, we also need more mathematics and theory, as some of the results presented here already call for. Thus, it is important to leave some space for thought experiments and excursions here, before which let us summarize what have been reaped and what should be the stratigic direction:

We have formulated the numerical RSF problem and showed that the precise analytical results of RSF established in Ref. 3 can be used to design effective semi-analytical algorithm for simulation. Explicit physical problems of the TGF and ABCF fashions have been created for proof-of-concept examples. The numerical results demonstrated the evolution of RSF distinct from that of NSF, the excitation of multi-scale RSF small eddies. Recent theoretical arguments and conjectures about the helicity effect on reducing the compressibility of the flow have been verified systematically, and actually extended in a firm way. A spectral transfer scenario is tested by various power spectra. The differences from different numerical discretization schemes are shown to be so small that they are irrelevant to the general conclusions about the production of small eddies and the helicity effects.

It is promising to further apply the algorithm to carry out the studies of RSF turbulence, forced or decaying, with other possibly more precise methods (such as the pseudo-spectra one, including the possibility of further improvement with the hexagonal Fourier transform[62]) and higher resolutions, and/or, longer-time simulations developing further from the results presented here or starting from random fields. Such studies could shed light on the passive scalar and rotating flow issues, for the obvious connections of RSF with the Taylor-Proudman limit. Note that, as the 'customized' strategy mentioned in Sec. II C would involve, the equation of RSF can be regarded as a completely new system with many of the spatial integrations 'canceling' those spatial derivatives in the orignal NSF. Such 'cancelation' between integration and derivative operators leads to the reduction of computational operations and presumably the corresponding numerical errors than those of NSF, contrary to the 'typical' strategy adopted here. Thus, there is still space of improvements on the algorithm, which could be helpful in high-resolution simulations and/or in high-precision analysis of specific flow properties of particular purposes beyond this note.

Further more, the formal compressible Taylor-Proudman theorem[4,10] requires the incompressibility of $\bm{u}_h$, thus it is also interesting to extend the algorithm to such flows and to carry out the numerical studies of such even more special RSFs, turbulent or not. As remarked in Ref. 3, it is not impossible that RSFs in $\mathbb{E}^d$ with $d > 3$ may also be of 'realistic' relevance. For example, the 2D and 3D passive scalar(s) may be regarded as, respectively, the $d-2$ or $d-3$ velocity component(s) independent on the corresponding coordinate(s). With some of the velocity gradients along some coordinates already being vanishing, the RSF or its appropriate variant is 'closer', than the hulking $d$-space full NSF, to the passive-scalar problem; and, RSF is enriched with finer thermodynamic and 'vortic' structures established in Ref. 3, thus presumably more powerful for sharper physical insights of the realistic passive scalar(s). Thus, in some sense, numerical studies of RSFs in general $d$-space, with the extension of our algorithm there, are also realistically desired.

On the other hand, although it appears in the case presented in Sec. IV B that the RSF with Stokesian viscosity taking the velocity dilatation into account, simulated with our semi-analytical algorithm, is also very close to that without the compressibility effect in the viscosity model, such a situation is not assured for every other case, especially for turbulence at very high Reynolds numbers and/or very large Mach numbers, and, for other flows (such as the non-newtonian fluid and quantum flows[63]) beyond the Navier-Stokes framework. Thus, other effective brute-force algorithms more universally applicable for general RSF simulations are also desired. Investigation of the influence of helicity on the complex singularities with the current viscosity model makes its own sense and should be of sufficient fundamental value, which still needs uniting high-fidelity schemes or methods so that the dissipation range is more accurate. The pseudo-spectral method (combined with our semi-analytical algorithm) may satisfy such a purpose but would require high resolution to resolve/capture the shocks and would need advanced techniques to detect the complex singularity structures.

Indeed, for the nonbarotropic RSF, a set of precise relations for the temperature structures can also be established[3] and be used to check the errors by measuring the deviations from them, just as we showed in Fig. 4 for the density structure. Preliminary numerical experiments show that our algorithm can also simulate such RSFs reasonably well: this is not like the situation for the 'zeroth order' algorithm discussed with that figure, because our semi-analytical algorithm already ensures the part of the thermodynamic structure and the difference of the temperature from our algorithm to the precise one in general moderate cases do not affect the flow too much (part of the reason why the isothermal process can be a good approximation, particularly when there is no heat input). If the RSF solution is unique and stable, particularly to the temperature perturbations, we do not seem to need extra constraint (otherwise 'over-determined'); but, the former being quite sure in the general compressible Navier-Stokes framework though, we see no guarantee for the latter. Thus, in principle, especially for turbulence at high $Re$ and $Ma$ where entropy modes are crucial and the fine structures are important for particular interests, or closely relevant problems involving thermal convection (e.g., Refs. 55 and 64), we should have a self-consistent algorithm, say, with such equations like (1a and 1b) that the relevant errors are completely under control, which deserves further studies.

To see how interesting and challenging, to our understanding, the open problem we set out is, let us elaborate a bit by taking some results from Ref. 3: For an ideal gas with, say, $p = \rho \mathcal{R} T$, Eq. (10) reads

$$\left[\frac{\nabla_h(\rho T)}{\rho}\right]_{,3} = 0. \tag{34}$$

Eq. (34) indicates that $\frac{\nabla_h(\rho T)}{\rho}$ is a function of only $x_1$ and $x_2$ and should have such separation of the variables

$$\rho = r(x_3)/R(x_1, x_2) \tag{35}$$

that the numerator and denominator can cancel the common factor $r(x_3)$. In other words, we have the same structure, especially Eq. (1a), as in the isothermal case. And, by taking (35) into (34), we further have

$$T(x_1, x_2, x_3) = \mathcal{T}(x_1, x_2) + R(x_1, x_2)\tau(x_3), \tag{36}$$

where $\mathcal{T}$ and $\tau$ and other variables are also time dependent. Eq. (36) characterizes the very fine structures of $T$, and designing an effective algorithm for it, semi-analytical or brute-force, appears to be a nontrivial challenge.

More definite derivatives from Eq. (36) have also been offered in Ref. 3, but, here, it is more appropriate to leave the space for imaginations. For example, instead of desperately looking for and exploiting even more precise and finner structures of $T$ following Ref. 3, we may deliberately use Eq. (36) as a variational constraint to derive the equation(s) appropriate for 'precise' (in the sense of keeping the fine structures pertaining to RSF) algorithm, which belongs to another special communication. However, we iterate that from the 'micro-physics' point of view, there exists "the possibility of appropriate 'gene insertion' at the 'micro-scopic' (molecular dynamics) and/or meso-scopic (kinetic theory) levels to obtain the 'macro-scopic' RSF".[3] For new algorithm(s) of, say, new models of real compressible fluids with realistic cubic equation of state (as a referee kindly suggested), our 'semi-analytical' numerical results here may serve as the 'benchmark' for testing.

Finally, we should emphasize that, being a first presentation of the numerical work on RSF, already a bit long though, many other important aspects can not be covered. For example, even for the deterministic acceleration scheme, its variations and effects on large scales deserve special investigations[51] (for most recent relevant discussions in other models, see, e.g., Refs. 54, 55, 65, and 66), and the elaborations of the connections with fundamentals of turbulence and physical realizations should be further carried out[24]: considering the domain of $2\pi$ periods, we have summarized the connection between the ACSL and the construction of most marked Jiu-Gong/Ba-Gua pattern of cyclones with $E_h$ concentrated at $k_h = \sqrt{2}$ for $s_a = 0$, and we have pointed out that the other extreme with $s_a = 1$ for $E_h$ concentrating at $k_h = 1$ does not present such a pattern resembling the cyclone cluster around Jupiter's northern polar. It is not our immediate ambition to reconcile the deep-convection and shallow-water ideas about Jupiter, but it does be important to further clarify the issue with the new relevant model and insights, and, to examine the vitality of RSFs, in both fundamentals and applications.[68]

## ACKNOWLEDGMENTS

Messrs H. Ren and C. Tang subsequently participated in some numerical experiments with other algorithms we tried in the early stage of the work, such as the 'zeroth order' scheme addressed in Sec III B, back in 2018[67] immediately after the flow was discovered[26]. We also thank Y. Zhang for sharing the well-structured code with different schemes for shock capturing, useful for the "standard-CFD part" of the RSF solver. T. Cai kindly imparted some relevant literatures in Ref. 68. This work was partially supported by the NSFC (Grant No. 11672102) and fully spiritualized by the Tian-Yuan-Xue-Pai foundation.

## DATA AVAILABILITY

The data that support the findings of this study are available from the corresponding author upon request.

low Mach and Rossby number limit, by D. Bresch, B. Desjardins, D. Gérard-Varet, "Rotating fluids in a cylinder," Discrete Contin. Dyn. Syst. 11, 47–82 (2004), and many follow-ups have studied various other limits, but not our 2C2Dcw1D3D RSF, of compressible rotating flows.

[5] K. Julien & E. Knobloch, "Reduced models for fluid flows with strong constraints," J. Math. Phys. 48, 165405 (2007).

[6] To our best knowledge, the first application of (real) Schur form of the velocity gradient matrix (VGM) with the attempt to distinguish rotation and shearing, among others, is due to Z. Li, X.-W. Zhang & F. He ["Evaluation of vortex criteria by virtue of the quadruple decomposition of velocity gradient tensor," Acta Phys. Sin. 60, 094702 (2014), with more detailed and explicit results of complex and real Shur transformations four years further back in Z. Li, "Theoretical Study on the Definition of Vortex," Master Degree Thesis (supervised by F. He), Qing-Hua University (2010)] who proposed the 'canonical rotation' part of the VGM as the most fundamental local criteria of vortex from the VGM 'normal' real Schur form, thus *Li et al.'s unique vortex* (Liutex). Various features of such Liutex, common to the 'swirling strength' [J. Zhou, R. J. Adrian, S. Balachandar, and T. M. Kendall, "Mechanisms for generating coherent packets of hairpin vortices in channel flow," J. Fluid Mech. 387, 353–396 (1999)], have recently been questioned by V. Kolář and J. Šistek ["Consequences of the close relation between Liutex and swirling strength," Phys. Fluids 32, 091702 (2020)], and the physics of fluids with VGM uniformly in the real Schur form with no complex eigenvalues, completely forbidding Liutex over space and time, are discussed in another communication. And, for other most recent applications, see, e.g., J.-L. Yu, Z.-Y. Zhao, and X.-Y. Lu, "Non-normal effect of the velocity gradient tensor and the relevant subgrid-scale model in compressible turbulent boundary layer," Physics of Fluids 33(2), 025103 (2021), and references therein. Studying the VGM dynamics with the coordinate transformations to specific frames has a long history [e.g., M. Carbone, M. Iovieno, and A. D. Bragg, "Symmetry transformation and dimensionality reduction of the anisotropic pressure Hessian," J. Fluid Mech. 900, A38 (2020) and references therein]. Now, the general real Schur frame is not unique, but we may impose some additional condition to fix it. Then, the specific real Schur frame is varying in space and time, introducing some complexity, but it is possible to be used to obtain analytical and physical insights of RSF or even NSF: P. Vieillefosse ["Local interaction between vorticity and shear in a perfect

tion, the quantum pressure term and the viscosity in the latter case do not admit the precise relation established in Ref. 3 and additional wisdoms are needed to compute the corresponding RSF: we particularly point this out because of the quantum vortex in rotating superfluid or Bose-Einstein condensate which appears to be closely relevant to the Taylor-Proudman limit of compressible flows, the latter being responsible for a particular horizontally-incompressible RSF[10]. For the less-known quantum Navier-Stokes, see, *e.g.*, most recently, J. W. Dong and Q. C. Ju, "Blow-up of smooth solutions to compressible quantum Navier-Stokes equations," Sci Sin Math 50, 873–884 (2020); B. L. Guo and B. Q. Xie, "Global existence of weak solutions to the three-dimensional full compressible quantum equations," Ann. of Appl. Math. 34, 1–31 (2018); J. W. Yang, G. H. Peng, H. Y. Hao and F. Z. Que, "Existence of global weak solution for quantum Navier-Stokes system," International Journal of Mathematics 31, 2050038 (2020); and references therein.

[64] Y. D. Afanasyev and Y.-C. Huang, "Poleward translation of vortices due to deep thermal convection on a rotating planet," Geophysical & Astrophysical Fluid Dynamics 114(6), 821–834 (2020).

[65] S. R. Brueshaber, K. M. Sayanagi and T. E. Dowling, "Dynamical regimes of giant planet polar vortices," Icarus 323, 46 (2019); S. R. Brueshaber and K. M. Sayanagi, "Effects of forcing scale and intensity on the emergence and maintenance of polar vortices on Saturn and Ice Giants," Icarus 361, 114386 (2021).

[66] F. Garcia, F. R. N. Chambers, A. L. Watts, "Deep model simulation of polar vortices in gas giant atmospheres," Monthly Notices of the Royal Astronomical Society 499(4), 4698–4715 (2020).

[67] H. Ren and J.-Z. Zhu, "Statistical topological fluid mechanics: kinetic and dynamic studies of flows with real Schur form velocity gradients." 10th National Conference of Fluid Mechanics. Hangzhou, Zhejiang, China (2018).

[68] Both Juno's photographs and our results indicate richer physics much beyond the state-of-the-art of the '$N+1$'-vortex analyses: e.g., J. J. Xue, E. R. Johnson & N. R. McDonald, "New families of vortex patch equilibria for the two-dimensional Euler equations," Phys. Fluids 29 (12), 123602 (2017) and J. N. Reinaud, "Three-dimensional quasi-geostrophic vortex equilibria with $m$-fold symmetry," J. Fluid Mech. 863, 32–59 (2019).